\documentclass[11pt,a4paper]{article}
\pdfoutput=1 

\usepackage[table]{xcolor}
\usepackage{colortbl}
\RequirePackage{ifpdf} 
\usepackage{amsmath} 
\usepackage{mathtools}
\usepackage{cases}

\usepackage{jheppub}
\usepackage{pstricks}
\usepackage[final]{pdfpages} 
\usepackage{ifpdf} 
\usepackage{slashed}

\usepackage{color}
\usepackage{xcolor}
\definecolor{urlblue}{rgb}{0.2,0.4,0.7}
\definecolor{citegreen}{rgb}{0,0.6,0.2}
\definecolor{linkred}{rgb}{0.9,0.2,0.1}
\usepackage{hyperref}
\hypersetup{
colorlinks=true, citecolor=citegreen, linkcolor=blue, urlcolor=urlblue}

\usepackage{graphics}
\usepackage{etoolbox} 
\usepackage{fixmath}
\usepackage{psfrag}

\usepackage{notoccite} 

\usepackage{amsfonts}
\usepackage{autobreak}
\usepackage{marginnote}

\usepackage{tikz}
\usetikzlibrary{positioning,arrows}
\usetikzlibrary{decorations.pathmorphing}
\usetikzlibrary{decorations.markings}
\usetikzlibrary{shapes.geometric}
\tikzset{
    vector/.style={decorate, decoration={snake}, draw},
    provector/.style={decorate, decoration={snake,amplitude=2.5pt}, draw},
    antivector/.style={decorate, decoration={snake,amplitude=-2.5pt}, draw},
    fermion/.style={draw=black, postaction={decorate},decoration={markings,mark=at position .55 with {\arrow[draw=black]{>}}}},
    fermionbar/.style={draw=black, postaction={decorate},
                       decoration={markings,mark=at position .55 with {\arrow[draw=black]{<}}}},
    fermionnoarrow/.style={draw=black},
    gluon/.style={decorate, draw=black,decoration={coil,amplitude=4pt, segment length=5pt}},
    scalar/.style={dashed,draw=black, postaction={decorate},decoration={markings,mark=at position .55 with {\arrow[draw=black]{>}}}},
    scalarbar/.style={dashed,draw=black, postaction={decorate},decoration={markings,mark=at position .55 with {\arrow[draw=black]{<}}}},
    scalarnoarrow/.style={dashed,draw=black},
    electron/.style={draw=black, postaction={decorate},decoration={markings,mark=at position .55 with {\arrow[draw=black]{>}}}},
    bigvector/.style={decorate, decoration={snake,amplitude=4pt}, draw},
}

\title{Polarised Amplitudes and Soft-Virtual Cross Sections for $b\bar b \rightarrow ZH$ at NNLO in QCD}

\author{Taushif Ahmed$^{a}$, A.~H. Ajjath$^{b}$, Long Chen$^{a}$, Prasanna K. Dhani$^{c}$,  Pooja Mukherjee$^{b}$ and V. Ravindran$^b$}
\emailAdd{taushif@mpp.mpg.de,
ajjathah@imsc.res.in,
longchen@mpp.mpg.de,
prasannakumar.dhani@fi.infn.it,
poojamukherjee@imsc.res.in,
ravindra@imsc.res.in}

\affiliation{$^a$Max-Planck-Institut f\"ur Physik, Werner-Heisenberg-Institut, 80805 M\"unchen, Germany \\
$^b$The Institute of Mathematical Sciences, HBNI, Taramani, Chennai 600113, India \\
$^c$INFN, Sezione di Firenze, I-50019 Sesto Fiorentino, Florence, Italy}

\preprint{IMSc/2019/10/09,MPP-2019-202}

\abstract{
Production of the Higgs boson, $H$ in association with a massive vector boson, $V$, i.e.,$~$the $VH$ process, plays an important  role in the explorations of Higgs physics at the Large Hadron Collider, both for a precise study of Higgs' Standard Model couplings and for probing New Physics. In this publication we present the two-loop corrections in massless quantum chromodynamics (QCD) to the amplitude of the Higgs production associated with a $Z$ boson via the bottom quark-antiquark annihilation channel with a non-vanishing  bottom-quark Yukawa coupling, which is a necessary ingredient of the full next-to-next-to-leading-order QCD corrections to the $VH$ process in the five-flavour scheme. The computation is performed by projecting the D-dimensional scattering amplitude directly onto an appropriate set of Lorentz structures related to the linear polarisation states of the $Z$ boson. We provide analytic expressions of the complete set of renormalised polarised amplitudes in terms of polylogarithms of maximum weight four. To give an estimation of the size of contributions from amplitudes considered in this work, we compute numerically the resulting cross sections under the soft-virtual approximation. We also take the opportunity to make a dedicated discussion regarding an interesting subtlety appearing in the conventional form factor decomposition of amplitudes involving axial currents regularised in D dimensions.
}

\begin{document}
\allowdisplaybreaks[4]
\unitlength1cm
\maketitle
\flushbottom


\def\D{{\cal D}}
\def\DD{\overline{\cal D}}
\def\g{\overline{\cal C}}
\def\gm{\gamma}
\def\M{{\cal M}}
\def\ep{\epsilon}
\def\epm1{\frac{1}{\epsilon}}
\def\epm2{\frac{1}{\epsilon^{2}}}
\def\epm3{\frac{1}{\epsilon^{3}}}
\def\epm4{\frac{1}{\epsilon^{4}}}
\def\unM{\hat{\cal M}}
\def\ashat{\hat{a}_{s}}
\def\asmur{a_{s}^{2}(\mu_{R}^{2})}
\def\sigbar{{{\overline {\sigma}}}\left(a_{s}(\mu_{R}^{2}), L\left(\mu_{R}^{2}, m_{H}^{2}\right)\right)}
\def\sigbarn{{{{\overline \sigma}}_{n}\left(a_{s}(\mu_{R}^{2}) L\left(\mu_{R}^{2}, m_{H}^{2}\right)\right)}}
\def\unas{ \left( \frac{\hat{a}_s}{\mu_0^{\epsilon}} S_{\epsilon} \right) }
\def\rnM{{\cal M}}
\def\bt{\beta}
\def\cD{{\cal D}}
\def\cC{{\cal C}}
\def\ca{\text{\tiny C}_\text{\tiny A}}
\def\cf{\text{\tiny C}_\text{\tiny F}}
\def\ct{{\red []}}
\def\sv{\text{SV}}
\def\murOmu{\left( \frac{\mu_{R}^{2}}{\mu^{2}} \right)}
\def\bb{b{\bar{b}}}
\def\bt0{\beta_{0}}
\def\bt1{\beta_{1}}
\def\bt2{\beta_{2}}
\def\bt3{\beta_{3}}
\def\gm0{\gamma_{0}}
\def\gm1{\gamma_{1}}
\def\gm2{\gamma_{2}}
\def\gm3{\gamma_{3}}
\def\nn{\nonumber}
\def\l{\left}
\def\r{\right}
\def\T{{\cal Z}}    
\def\U{{\cal Y}}

\def\nn{\nonumber\\}
\def\ep{\epsilon}
\def\T{\mathcal{T}}
\def\V{\mathcal{V}}

\newcommand\myeq{\stackrel{\mathclap{\normalfont\mbox{\tiny FR}}}{=}}

\def\qgraf{{\fontfamily{qcr}\selectfont
QGRAF}}
\def\python{{\fontfamily{qcr}\selectfont
PYTHON}}
\def\form{{\fontfamily{qcr}\selectfont
FORM}}
\def\reduze{{\fontfamily{qcr}\selectfont
REDUZE2}}
\def\kira{{\fontfamily{qcr}\selectfont
Kira}}
\def\litered{{\fontfamily{qcr}\selectfont
LiteRed}}
\def\fire{{\fontfamily{qcr}\selectfont
FIRE5}}
\def\air{{\fontfamily{qcr}\selectfont
AIR}}
\def\mint{{\fontfamily{qcr}\selectfont
Mint}}
\def\hepforge{{\fontfamily{qcr}\selectfont
HepForge}}
\def\arXiv{{\fontfamily{qcr}\selectfont
arXiv}}
\def\Python{{\fontfamily{qcr}\selectfont
Python}}
\def\ginac{{\fontfamily{qcr}\selectfont
GiNaC}}
\def\polylogtools{{\fontfamily{qcr}\selectfont
PolyLogTools}}
\def\anci{{\fontfamily{qcr}\selectfont
Finite\_ppbk.m}}
\def\gosam{{\fontfamily{qcr}\selectfont
GoSam}}
\def\fermat{{\fontfamily{qcr}\selectfont
fermat}}
\def\xml{{\fontfamily{qcr}\selectfont
qgraf-xml-drawer}}

\newcommand{\dis}{}
\newcommand{\overbar}[1]{mkern-1.5mu\overline{\mkern-1.5mu#1\mkern-1.5mu}\mkern
1.5mu}
\newcommand{\TODO}[1]{ {\color{red} #1} }

\section{Introduction}
\label{sec:intro}

Ever since the discovery of the 125 GeV Higgs boson at the Large Hadron Collider~\cite{Aad:2012tfa,Chatrchyan:2012xdj} (LHC), the detailed investigation of its dynamical properties, i.e.,~how it interacts with (other) known fundamental particles, remains among the major research topics of the current and future particle physics programs. 
The interactions of this Higgs boson explored so far are in accord with the predictions of the Standard Model (SM), and considerable improvement of the experimental precision is expected with the future high luminosity LHC program. 
(See, for instance, ref.~\cite{Brandstetter:2018eju} and references therein for a brief overview.)

One of the recent achievements regarding the investigation of the 125 GeV Higgs boson's couplings to fermions was the direct observation of its decay to a pair of bottom quarks by the ATLAS and CMS experiments~\cite{Aaboud:2018zhk,Sirunyan:2018kst}. 
The main contributing channels to this important experimental result are from production processes in which the Higgs boson is produced in association with a $W$ or $Z$ boson, known as the $VH$ process, and the associated electroweak vector boson is typically chosen to be reconstructed via its leptonic decays. 
The presence of the vector boson in the final state in addition to the Higgs boson is crucial to substantially reduce the SM backgrounds, for instance by requiring a large transverse momentum of the associated vector boson~\cite{Butterworth:2008iy}.
Additional selection criteria are also imposed to enrich the signal $VH$ events over backgrounds eventually to a manageable level to allow for such an experimental observation~\cite{Aaboud:2018zhk, Sirunyan:2018kst}.
Accordingly, on the theoretical side, it is thus very desirable to have a precise knowledge about the $VH$ process at hadron colliders, especially to meet the foreseeable precision requirements from future experiments for studies of the 125 GeV Higgs boson (as well as potentially non-standard Higgs bosons) with ever more details.~\\

\begin{figure}[htbp]
\begin{center}
\includegraphics[scale=0.38]{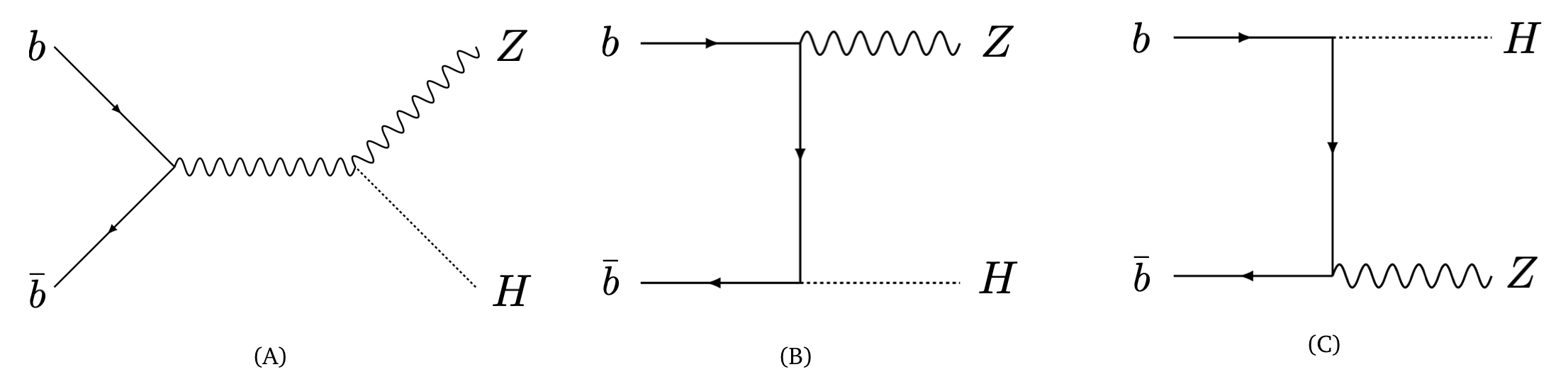}
\caption{Feynman diagrams at leading order}
\label{dia:tree}
\end{center}
\end{figure}

Given the aforementioned phenomenological importance of $VH$ productions, there have been many computations available in the literature on this subject aiming to improve theoretical predictions as much as possible. 
Main production channels at the LHC include the quark-induced and gluon-induced $VH$ processes.
The focus of this article is a part of the $b$-quark-induced $ZH$ process that involves a non-vanishing Yukawa coupling $\lambda_b$  between the $b$ quark and the Higgs boson. 
At the tree level there are three contributing diagrams for the $b$-quark-induced $ZH$ process\footnote{This holds in the physical unitary gauge where unphysical degree-of-freedoms in electroweak ghosts and would-be-Goldstone bosons decouple from the spectrum.}, as shown in figure~\ref{dia:tree}.
The diagram (A) gives an $s$-channel contribution with the same structure as that of the Drell-Yan production, which has been studied extensively in QCD up to order $\mathcal{O}(\alpha_s^2)$ in refs.~\cite{Altarelli:1978id,Han:1991ia,Baer:1992vx,Ohnemus:1992bd, Mrenna:1997wp,Brein:2003wg,Ferrera:2014lca,Campbell:2016jau,Ferrera:2017zex} and to $\mathcal{O}(\alpha_s^3)$ in refs.~\cite{Ahmed:2014cla,Li:2014bfa,Catani:2014uta,Kumar:2014uwa}, also at lepton colliders, e.g.~in \cite{Barger:1993wt,Blumlein:2019pqb}.
The presence of this contribution is independent of $\lambda_b$ as the Higgs boson is radiated by bremsstrahlung of the $Z$ boson, which makes this channel particularly valuable for studying $VVH$ vertex. In diagrams (B) and (C) the Higgs boson is coupled directly to the $b$ quark via its Yukawa interaction. Since the $Z$ boson appears in this non-Drell-Yan type diagrams only as an external on-shell leg, there is no explicit gauge-fixing parameter involved and this part of the contributions is manifestly gauge invariant. If one keeps the $b$ quark kinematically massless ($m_b = 0$), these two types of contributions do not interfere and hence can be treated separately. This is because the \textit{chirality} of the $b$ quark line is preserved in diagram (A) but flipped exactly once in diagrams (B) and (C).

Starting from $\mathcal{O}(\alpha_s^2)$ in QCD a class of two-loop diagrams appears contributing to the $b$-quark initiated $ZH$ process in which the Higgs couples directly to a closed  top-quark loop. This was studied, e.g.~, in refs.~\cite{Brein:2003wg,Brein:2011vx} by making use of asymptotic expansions in the heavy-top limit. In addition, at order $\alpha_s^2$ the gluon-fusion induced $ZH$ production also opens up, which actually contributes considerably due to the rather large gluon luminosity at the LHC~\cite{Kniehl:1990iva,Brein:2011vx,Altenkamp:2012sx}. 
While these Drell-Yan type contributions and top-quark loop induced corrections (in the heavy-top limit) have already been studied extensively in the literature, all of which are independent of $\lambda_b$, the work presented in this article  focuses on the  order $\alpha_s^2$ QCD corrections
 to the leading-order diagrams (B) and (C). We keep the $b$ quark kinematically massless in our analytic computation of two-loop amplitudes involved.  Keeping $\lambda_b \neq 0$ while $m_b = 0$ amounts to retaining just the leading contribution of the  result with full $b$ quark mass dependence, an approximation similar to what was adopted in a series of impressive works on $H \rightarrow b\bar{b}$~\cite{Chetyrkin:1996sr,Kataev:1997cq,Chetyrkin:1997vj,Chetyrkin:1997iv,Baikov:2005rw,Ravindran:2006bu,Ravindran:2006cg,Gehrmann:2014vha,Ahmed:2014cha,Ahmed:2014era,Davies:2017xsp,Herzog:2017dtz,Mondini:2019gid,Duhr:2019kwi,H:2019nsw,H:2019dcl}\footnote{We remark that the full $b$-quark mass dependence of $H \rightarrow b\bar{b}$ was studied at order $\alpha_s^2$, e.g.~at the inclusive level in refs.~\cite{Kataev:1993be,Chetyrkin:1997mb} and at the differential level in ref.~\cite{Bernreuther:2018ynm}.}. Recently, by keeping the non-zero bottom Yukawa coupling, the di-Higgs production cross section at next-to-next-to-leading order (NNLO) in QCD at threshold is also achieved in ref.~\cite{H:2018hqz}.

In the SM, quarks acquire their masses through Yukawa interactions with one Higgs doublet which develops a non-vanishing vacuum expectation value (\textit{vev}) leading to electroweak symmetry breaking. Consequently the quark Yukawa coupling strengths are simply proportional to their respective masses divided by \textit{vev}. However, in certain beyond SM scenarios with more than one Higgs doublet, 
e.g.,~the Minimal Supersymmetric SM~\cite{Nilles:1983ge} (MSSM), the bottom-quark Yukawa coupling can be enhanced w.r.t. the top-quark Yukawa coupling in the large $\mathrm{tan}\beta$ region, where $\mathrm{tan}\beta$ is the ratio of \textit{vev}s of up- and down-type Higgs fields in the Higgs sector of the MSSM. Such scenarios motivate the detailed investigation of the $b$-quark-induced $ZH$ process that is proportional to $\lambda_b$.  
Though the contribution and consequently, its impact on the LHC is expected to be small because of the small $\lambda_b$, nevertheless to make a quantitative analysis of the contribution from these channels at NNLO we compute numerically their cross sections under the soft-virtual approximation~\cite{Ravindran:2005vv,Ravindran:2006cg} in the five-flavour scheme.~\\

The major motivation for the work considered in this article arises from some interesting questions in the computation of the non-Drell-Yan type  $b\bar{b} \rightarrow ZH$ scattering amplitude at higher orders in QCD, which may be potentially relevant in a larger scenario.
We address subtleties appearing in the conventional form factor decomposition of loop amplitudes involving an axial current in D dimensions (with a non-anticommuting prescription for $\gamma_5$): whether we need to include all evanescent Lorentz structures\footnote{By ``evanescent Lorentz structures'' we refer to linear structures in a Lorentz tensor decomposition that are non-vanishing in D-dimensions but vanishing in 4 dimensions, whose presence is a consequence of the D-dimensional complete basis being larger than that in 4 dimensions. (See, for instance, ref.~\cite{Chen:2019wyb} for more detailed discussions.)} to end up with correct results in computations made in D-dimensions; whether the particular regularisation prescription implied by projectors prescribed recently in ref.~\cite{Chen:2019wyb} remains unitary at higher orders once applied to this scattering process, etc. The two-loop correction to the non-Drell-Yan type diagrams in figure~\ref{dia:tree} provides a non-trivial case to investigate these issues. 
One of the main messages conveyed through this article is that even if the loop amplitudes are not defined or regularised strictly in the 't Hooft-Veltman scheme~\cite{tHooft:1972tcz}, expressions for projectors derived in four dimensions are still sufficient and lead to correct results (for physical observables), be the projectors corresponding to conventional form factors or to polarise amplitudes of the process considered. This is particularly helpful in evaluating loop amplitudes for cases involving axial currents (if a non-anticommuting $\gamma_5$ should be used).
We will discuss these questions in detail through the work presented in this article.
~\\

The article is organised as follows.
In the next section, we set up the kinematics and specify a few basic aspects of the computation. 
In section~\ref{sec:projectors} prescriptions used for constructing projectors are exposed in detail. 
In particular, the subsection~\ref{sec:projectors_LP} is devoted to a self-contained discussion of a set of  projectors that correspond  directly to amplitudes in the 
linear polarisation basis. The Lorentz structures needed in the conventional form factor decomposition and the corresponding projectors are given in 
subsection~\ref{sec:projectors_FF} with a dedicated discussion of an interesting subtlety appearing in the conventional form factor decomposition 
of amplitudes involving axial currents regularised in D dimensions.
Section~\ref{sec:uv} gives details of the ultraviolet renormalisation prescription we adopted as well as that of the verification 
of the axial quantum anomaly relation in our computational setup. 
In section~\ref{sec:computation}, we outline the flowchart, especially the tool chain, employed to accomplish the computation of amplitudes presented through this work. 
We verify in section~\ref{sec:irfr} the universal infrared divergences appearing in our ultraviolet renormalised amplitudes, with emphasis on the 
regularisation-scheme independence of the four-dimensional finite remainders obtained in this way. In particular, we address the aforementioned questions.
To estimate the size of the contributions arising from channels considered in this work we compute numerically the cross sections in the soft-virtual 
approximation in section~\ref{sec:cssv}.
We conclude in section \ref{sec:con}.

\section{Preliminaries}
\label{sec:setup}

We consider the production of a scalar Higgs boson, $H$, in association with a massive vector boson, $Z$, through bottom quark anti-quark annihilation
\begin{align}
\label{eq:process}
    b(p_1) + \bar{b}(p_2) \to Z(q_1) + H(q_2)\,.
\end{align}
Here $b(\bar b)$ denotes the bottom quark (anti-quark). The quantity within the parenthesis represents the momentum of the corresponding particle satisfying on-shell conditions $p_i^2=0,q_1^2=m_z^2,q_2^2=m_h^2$, where $m_z$ and $m_h$ are the mass of the $Z$ and Higgs boson, respectively.  The Mandelstam variables are defined as 
\begin{align}
\label{eq:Mandelstam}
    s\equiv (p_1+p_2)^2\,,~~ t\equiv (p_1-q_1)^2\, ~~\text{and}~~ u\equiv (p_2-q_1)^2
\end{align}
satisfying $s+t+u=q_1^2+q_2^2=m_z^2+m_h^2$. 
The physical region of the phase space is bounded by $t \, u=q_1^2 \, q_2^2$ such that it satisfies
\begin{align}
    s \ge \left( \sqrt{q_1^2}+\sqrt{q_2^2}\right)^2\,, ~~~~ \frac{1}{2} \left(q_1^2+q_2^2-s-\kappa\right) \le t \le \frac{1}{2} \left(q_1^2+q_2^2-s+\kappa\right),
\end{align}
where $\kappa$ is the K\"all\'en function defined by 
\begin{align}
    \kappa(s,q_1^2,q_2^2)\equiv \sqrt{s^2+q_1^4+q_2^4-2(s q_1^2+q_1^2 q_2^2+s q_2^2)}\,.
\end{align} 

As already mentioned in the introduction, we keep a non-zero Yukawa coupling but only for the $b$ quark, which otherwise is set to be kinematically massless. We compute through this article only QCD corrections to the non-Drell-Yan type diagrams in figure~\ref{dia:tree} that depends on $\lambda_b$, in  $n_f=5$ flavour massless QCD. 
The $Z$-boson interacts with all massless quarks through the respective vector and axial couplings. Any multi-loop calculation involving axial coupling in dimensional regularisation~\cite{tHooft:1972tcz,Bollini:1972ui} (DR) faces the problem of defining the inherently 4-dimensional objects, Dirac's $\gamma_5$ (and Levi-Civita symbol $\epsilon^{\mu\nu\rho\sigma}$), properly in D-dimensions. In this article, we follow the definition of $\gamma_5$ in dimensional regularisation which was introduced by 't Hooft-Veltman~\cite{tHooft:1972tcz} and Breitenlohner-Maison~\cite{Breitenlohner:1977hr}
\begin{align}
\label{eq:gamma5}
	\gamma_5=\frac{i}{4!}\varepsilon_{\mu\nu\rho\sigma}\gamma^{\mu}\gamma^{\nu}\gamma^{\rho}\gamma^{\sigma}\,.
\end{align}
However, the $\gamma_5$ defined through the above equation no longer fully anti-commutes with the D-dimensional $\gamma^{\mu}$, which has profound consequences in computations involving axial currents in D-dimensions. In particular, the aforementioned definition of $\gamma_5$ affects the ultraviolet renormalisation non-trivially~\cite{Chanowitz:1979zu,Trueman:1979en,Larin:1991tj,Larin:1993tq}, to be addressed in details in section~\ref{sec:uv}. In order to define the Hermitian axial current correctly we need to symmetrise it~\cite{Akyeampong:1973xi,Larin:1991tj} before using the definition \eqref{eq:gamma5}
\begin{align}
\label{eq:axial-symm}
	\gamma_{\mu}\gamma_5 \rightarrow \frac{1}{2} \Big(\gamma_{\mu}\gamma_5-\gamma_5\gamma_{\mu}\Big)\,.
\end{align}
By combining \eqref{eq:gamma5} and \eqref{eq:axial-symm}, we obtain~\cite{Akyeampong:1973xi,Larin:1991tj}
\begin{align}
\label{eq:gamma5-axial}
	\gamma_{\mu}\gamma_5 = \frac{i}{6} \varepsilon_{\mu\nu\rho\sigma} \gamma^{\nu} \gamma^{\rho} \gamma^{\sigma}
\end{align}
which is used in D-dimensions for our calculation. The contraction of pairs of Levi-Civita symbols appearing in the calculation is made through
\begin{align}
\label{eq:epsilon-contract}
	\varepsilon_{\mu_1\nu_1\rho_1\sigma_1} \varepsilon^{\mu_2\nu_2\rho_2\sigma_2}=4! \delta^{\mu_2}_{[\mu_1}\ldots\delta^{\sigma_2}_{\sigma_1]}
\end{align}
where all the indices carried by space-time metric tensors on the right hand side are (by definition) considered in D dimensions~\cite{Zijlstra:1992kj}. The symbol [\;] around the indices represents the anti-symmetric combination. In the next section, we discuss the projector method that is adopted to compute helicity amplitudes of the scattering process \eqref{eq:process}.

\section{The Prescription of Projectors}
\label{sec:projectors}

In this work helicity amplitudes of the process \eqref{eq:process} are obtained by projecting the D-dimensional amplitude in its defining Feynman-diagrammatic representation directly onto a minimal set of D-dimensional projection operators following the approach proposed in~\cite{Chen:2019wyb}. 
Since there has not been much discussion of this approach in the literature, we will first provide a quick recap of the essentials involved, while the reader is referred to ref.~\cite{Chen:2019wyb} for more details.

\subsection{Projectors for Linearly Polarised $b \bar{b} Z H$ Amplitudes}
\label{sec:projectors_LP}

Polarised amplitudes carry bookkeeping external polarisation state vectors. What typically stops one from viewing the product of these state vectors as a projector defined in the usual sense is simply the fact that these external state vectors are not explicitly given solely in terms of external momenta and/or algebraic constants. These are essentially the defining criteria of the usual projectors in order to end up with projections that are Lorentz invariant and dependent only on external kinematics. The polarisation projectors as prescribed in~\cite{Chen:2019wyb} are based on the momentum basis representations of external state vectors, and all their open Lorentz indices are by definition taken to be D-dimensional to facilitate a uniform projection with just one dimensionality D=$g_{~\mu}^{\mu}$. Essentially, the momentum basis representations of polarisation state vectors allow us to find a Lorentz covariant representation of the tensor products of external particles' state vectors (for both bosons and fermions, massless or massive) solely in terms of external momenta and algebraic constants, such as the metric tensor, Levi-Civita symbol and Dirac matrices.
As all their open Lorentz indices are promoted to be D-dimensional, no dimensional splitting is ever introduced for loop momenta and/or Lorentz indices of inner vector fields. In this way, the contraction of Lorentz indices is made commutable with loop integration. It is also owing to this fact that despite being different from the conventional dimensional regularisation scheme~\cite{Collins:1984xc} (CDR), all UV renormalisation constants and integrated IR subtraction coefficients needed to complete a full fixed order computation of polarised physical observables can be directly recycled from results obtained within CDR. The bare (helicity) amplitudes resulting from this computational scheme are in general different from those defined in main-stream regularisation schemes, such as CDR~\cite{Collins:1984xc}, 't Hooft-Veltman (HV)~\cite{tHooft:1972tcz}, Dimensional-Reduction (DRED)~\cite{Siegel:1979wq,Capper:1979ns,Jack:1993ws}, and  Four-Dimensional-Helicity (FDH) schemes~\cite{Bern:1991aq,Bern:2002zk}, while the properly defined finite remainders will of course be the same\footnote{This simply follows from this particular regularisation prescription being \textit{unitary}, to be discussed in section~\ref{sec:irfr} in more detail.}. 
All these points will be verified explicitly for the scattering amplitudes of \eqref{eq:process} up to two-loop order presented in this publication, in particular in section~\ref{sec:irfr}.

There are a number of reasons responsible for generating differences in the bare amplitudes as regularised in this way from their counterparts defined in the aforementioned main-stream dimensional regularisation schemes, as explained in ref.~\cite{Chen:2019wyb}. In particular, in our scheme the number of polarisation degrees-of-freedom (d.o.f) of an external particle is always equal to the number of its physical d.o.f in 4 dimensions while, however, all open Lorentz indices are by definition set to be D-dimensional to facilitate technically a uniform projection in D dimensions. Notably, all bookkeeping Levi-Civita symbols ubiquitously appearing in the momentum basis representations of certain transversal polarisation states are by definition manipulated according to \eqref{eq:epsilon-contract} with all resulting space-time metric tensors set to be D-dimensional. Notice that in both construction and application of these D-dimensional projectors onto amplitudes, there is no need to appeal to their Lorentz tensor decomposition representations first (and hence there is no question of whether or not the amplitude reconstructed from form factor decomposition is faithful in D dimensions).
~\\

The amplitude of \eqref{eq:process}, being multi-linear in the state vectors of the external particles to all loop orders in perturbative calculations, can be schematically parameterised as 
\begin{align} 
\label{eq:bbZHamplitude}
    \mathcal{M} &= \bar{v}(p_2) \, \mathbf{\Gamma}^{\mu} \, u(p_1) \, \varepsilon^{*}_{\mu}(q_1) \nonumber\\
    &= \bar{v}(p_2) \, \mathbf{\Gamma}^{\mu}_{vec} \, u(p_1) \, \varepsilon^{*}_{\mu}(q_1) 
    \,+\,\bar{v}(p_2) \, \mathbf{\Gamma}^{\mu}_{axi} \, u(p_1) \, \varepsilon^{*}_{\mu}(q_1)\nonumber\\
    &\equiv {\cal M}_{vec}+{\cal M}_{axi}\,.
\end{align}
The symbol $\mathbf{\Gamma}^{\mu} \equiv \mathbf{\Gamma}^{\mu}_{vec} + \mathbf{\Gamma}^{\mu}_{axi}$ denotes a matrix in the Dirac spinor space with one open Lorentz index $\mu$ which may be carried by either the elementary Dirac matrix $\gamma^{\mu}$ or one of the external momenta involved (such as $q_1^{\mu}$). 
It consists of contributions from the vector and axial coupling of the $Z$ boson, denoted respectively by 
$\mathbf{\Gamma}^{\mu}_{vec}$ and $\mathbf{\Gamma}^{\mu}_{axi}$.
As emphasized in the introduction,  this work deals with the non-Drell-Yan type contributions to \eqref{eq:process} in massless QCD where a non-vanishing Yukawa coupling $\lambda_b$ is retained. Due to the Yukawa coupling vertex on the external $b$ quark line, all non-vanishing amplitudes of this gauge invariant class of diagrams involve two external massless spinors, $u(p_1)$ and $v(p_2)$, with opposite \textit{chirality}. As a consequence, the power of elementary Dirac matrices in $\mathbf{\Gamma}^{\mu}$ sandwiched between the two external massless $b$ quark spinors must be even in order to have a non-vanishing matrix element between such a pair of spinors (with opposite chirality).

According to ref.~\cite{Chen:2019wyb}, we construct the following list of projectors, which in their bookkeeping forms read as 
\begin{align} 
\label{eq:LPprojectors_primitive}
    \bar{u}(p_1) \, {\mathbf{N}}_i \, v(p_2) \, \varepsilon^{\mu}_j \, , 
    \quad \text{for $i = s, p$ and $j = T,Y,L$} 
\end{align}
where the ${\mathbf{N}}_s = \mathbf{1}$, ${\mathbf{N}}_p = \gamma_5$, and $\varepsilon^{\mu}_j$ with $j = T,Y,L$ denote the three linear polarisation eigenstates of $Z(q_1)$ identified in the center-of-mass reference frame of the collision. To be more specific about this, we choose $\varepsilon^{\mu}_T$ to be the transversal polarisation within the scattering plane determined by the three linearly independent external momenta, $p_1, p_2, q_1$. The other transversal polarisation $\varepsilon^{\mu}_Y$ 
is orthogonal to $p_1, p_2,$ and $ q_1$, and is constructed using the Levi-Civita symbol. 
The third physical polarisation state of the $Z$ boson, its longitudinal polarisation denoted by the vector $\varepsilon^{\mu}_L$, has its spatial part aligned with its own momentum $q_1$.

The momentum basis representations of $\varepsilon^{\mu}_j$ thus defined in terms of $p_1, p_2, q_1$ can be determined in the following way. We first write down a Lorentz covariant parametrisation ansatz for the $\varepsilon^{\mu}_j$ and then solve the orthogonality and normalisation conditions of linear polarisation state vectors for the linear decomposition coefficients. Once we have established a definite Lorentz covariant decomposition form in 4 dimensions solely in terms of external momenta and kinematic invariants, this form will be used as the \textit{definition} of the corresponding polarisation state vector in D dimensions. Following this line, it is rather straightforward to arrive at the following result:
\begin{align} 
\label{eq:MBR_Zpolvec}
    \varepsilon^{\mu}_T &=  \mathcal{N}^{\, -1}_T\, 
    \Big(
    -\left(2 m_z^4 + u (t + u) - m_z^2 (2 s + t + 3 u)\right) \, p_1^{\mu} \nonumber\\
    & ~~~~~~ \, +  \left(2 m_z^4 + t (t + u) - m_z^2 (2 s + 3 t + u)\right) \, p_2^{\mu} \, + s (t - u) \, q_1^{\mu} 
    \Big) \, , \nonumber\\
    \varepsilon^{\mu}_Y &= \mathcal{N}^{\, -1}_Y \, 
    \Big( - \epsilon^{\mu \nu \rho \sigma} p_{1\, \nu} p_{2\, \rho} q_{1\, \sigma} \Big)  \, , \nonumber\\
    \varepsilon^{\mu}_L &= \mathcal{N}^{\, -1}_L \, 
    \Big( (2 m_z^2 - t - u)\, q_1^{\mu} - 2 m_z^2 (p_1^{\mu} + p_2^{\mu}) \Big)  \, ,
\end{align}
where the squares of the normalisation factors are 
\begin{eqnarray} \label{eq:ZpolvecNSQ}
\mathcal{N}_T^{\, 2} &=&  -s \left(4 m_z^4 + (t + u)^2 - 4 m_z^2 (s + t + u)\right) 
\left( m_z^4 + t u - m_z^2 (s + t + u)\right)  \, , \nonumber\\
\mathcal{N}_Y^{\, 2} &=& \frac{1}{4} s \left(-m_z^4 - t u + m_z^2 (s + t + u)\right)  \, , \nonumber\\
\mathcal{N}_L^{\, 2} &=& -m_z^2 \left(4 m_z^4 + (t + u)^2 - 4 m_z^2 (s + t + u)\right).
\end{eqnarray}
We start with the primitive bookkeeping form of the projectors in \eqref{eq:LPprojectors_primitive}, and then substitute \eqref{eq:MBR_Zpolvec} and subsequently simplify the expressions using 4-dimensional Lorentz and Dirac algebra (and 4-dimensional equations of motion) as much as possible.
In this way one ends up with one definite form of projectors to be used in D-dimensional calculations 
(see discussion in the beginning of section~\ref{sec:projectors_FF_axi}).
Eventually, after substituting the non-anticommutating $\gamma_5$ prescription, i.e.~\eqref{eq:gamma5} and \eqref{eq:gamma5-axial}, 
we \textit{choose} the following reduced definite form of the 6 projectors to be used in the projections of \eqref{eq:bbZHamplitude} according to the D-dimensional algebra:
\begin{align} 
\label{eq:LPprojectors_canonical}
    \mathcal{P}_1^{\mu} &= \bar{u}(p_1) \, v(p_2) \, \Big(
    -\left(2 m_z^4 + u (t + u) - m_z^2 (2 s + t + 3 u)\right) \, p_1^{\mu} \nonumber\\
    & ~~~~~~~~~~~~~~~~ \, +  \left(2 m_z^4 + t (t + u) - m_z^2 (2 s + 3 t + u)\right) \, p_2^{\mu} \, + s (t - u) \, q_1^{\mu} 
    \Big)\,,\nonumber\\
    \mathcal{P}_2^{\mu} &= \bar{u}(p_1) \, v(p_2) \, \Big( - \epsilon^{\mu \nu \rho \sigma} p_{1\, \nu} p_{2\, \rho} q_{1\, \sigma} \Big)\,, \nonumber\\
    \mathcal{P}_3^{\mu} &= \bar{u}(p_1) \, v(p_2) \, \Big( (2 m_z^2 - t - u)\, q_1^{\mu} - 2 m_z^2 (p_1^{\mu} + p_2^{\mu}) \Big)\,, \nonumber\\
    \mathcal{P}_4^{\mu} &= \bar{u}(p_1) \mathbold{\epsilon}_{\gamma \gamma \gamma \gamma} v(p_2) \, \Big(
    -\left(2 m_z^4 + u (t + u) - m_z^2 (2 s + t + 3 u)\right) \, p_1^{\mu} \nonumber\\
    & ~~~~~~~~~~~~~~~~ \, +  \left(2 m_z^4 + t (t + u) - m_z^2 (2 s + 3 t + u)\right) \, p_2^{\mu} \, + s (t - u) \, q_1^{\mu} 
    \Big)\,,\nonumber\\
    \mathcal{P}_5^{\mu} &= \bar{u}(p_1) \frac{i}{8} \Big((-2 m_z^2 + t + u) \Big( \slashed{p}_2 \gamma_{\mu} + \gamma_{\mu} \slashed{p}_1 \Big) + 2 s \Big( \gamma_{\mu} \slashed{q}_1 - \slashed{q}_1 \gamma_{\mu} \Big) \nonumber\\
    &~~~~~~~~~~~~+ 2 (u-t) \Big(p_{1\mu}+p_{2\mu}\Big) \Big) v(p_2)\,,\nonumber\\
    \mathcal{P}_6^{\mu} &= \bar{u}(p_1) \mathbold{\epsilon}_{\gamma \gamma \gamma \gamma} v(p_2) \, \Big( (2 m_z^2 - t - u)\, q_1^{\mu} - 2 m_z^2 (p_1^{\mu} + p_2^{\mu}) \Big)\, ,
\end{align}
where we have pulled out the normalisation factors as defined in \eqref{eq:ZpolvecNSQ} and the symbol $\mathbold{\epsilon}_{\gamma \gamma \gamma \gamma} \equiv -\frac{i}{24} \epsilon_{\mu \nu \rho \sigma} \gamma^{\mu} \gamma^{\nu} \gamma^{\rho} \gamma^{\sigma}$ is introduced. Upon pulling out the respective normalisation factors, all 6 projectors thus constructed have only a polynomial dependence on the kinematic variables and external momenta, which is very advantageous in practice. Computations of the projection of the amplitude \eqref{eq:bbZHamplitude} onto each of these 6 structures proceed in the usual way: the product of Dirac spinors will be cast into a trace of products of Dirac matrices with the aid of (unpolarised) Landau density matrices of external spinors. 
The contraction of pairs of Levi-Civita symbols appearing in the projection is made in accordance with \eqref{eq:epsilon-contract} 
where all indices of the resulting metric tensors are set to be D-dimensional by definition.
Furthermore, since by construction the momentum basis representations of polarisation state vectors in \eqref{eq:MBR_Zpolvec} fulfill all the defining physical constraints, the index contraction between these projectors \eqref{eq:LPprojectors_canonical} and the amplitude \eqref{eq:bbZHamplitude} is therefore always simply done with the space-time metric tensor $g_{\mu\nu}$.
Since the process \eqref{eq:process} in question has only 3 linearly independent external momenta, upon using total momentum conservation, there will be no term that involves explicitly the Levi-Civita tensor after the projections have been made using \eqref{eq:LPprojectors_canonical}.
It is thus straightforward to see that the projections with $\mathcal{P}_1^{\mu}$, $\mathcal{P}_3^{\mu}$, and $\mathcal{P}_5^{\mu}$ receive contributions only from the vector part of the $Z$ boson coupling to quark, while contributions from the $Z$ boson's axial coupling to quark survives only in the projections with $\mathcal{P}_2^{\mu}$, $\mathcal{P}_4^{\mu}$, and $\mathcal{P}_6^{\mu}$. In this way, the axial and vector part of the amplitude \eqref{eq:bbZHamplitude} get separated naturally.
~\\

Polarised amplitudes of this scattering process are thus first extracted in the linear polarisation basis using the 6 projectors as described above, from which helicity amplitudes can be readily composed afterwards. Helicity eigenstates of a massless spinor coincide with its chiral eigenstates,  and according to the usual convention, a left chiral massless $u$($v$)-type spinor has a negative (positive) helicity. 
Therefore in the end we compose 6 polarised amplitudes by the following linear combinations:
\begin{align} 
\label{eq:HLamps}
    \mathcal{M}^{[++T]} &= \mathcal{N}_{++T}^{-1} \, 
    \Big( \mathcal{P}^{\mu}_1 \,-\,  \mathcal{P}^{\mu}_4 \Big)\, \bar{v}(p_2) \, \mathbf{\Gamma}_{\mu} \, u(p_1) \, , \nonumber\\
    \mathcal{M}^{[++Y]} &= \mathcal{N}_{++Y}^{-1} \, 
    \Big( \mathcal{P}^{\mu}_2 \,-\,  \mathcal{P}^{\mu}_5 \Big)\, \bar{v}(p_2) \, \mathbf{\Gamma}_{\mu} \, u(p_1) \, , \nonumber\\
    \mathcal{M}^{[++L]} &= \mathcal{N}_{++L}^{-1} \, 
    \Big( \mathcal{P}^{\mu}_3 \,-\,  \mathcal{P}^{\mu}_6 \Big)\, \bar{v}(p_2) \, \mathbf{\Gamma}_{\mu} \, u(p_1) \, , \nonumber\\
    \mathcal{M}^{[--T]} &= \mathcal{N}_{--T}^{-1} \, 
    \Big( \mathcal{P}^{\mu}_1 \,+\,  \mathcal{P}^{\mu}_4 \Big)\, \bar{v}(p_2) \, \mathbf{\Gamma}_{\mu} \, u(p_1) \, , \nonumber\\
    \mathcal{M}^{[--Y]} &= \mathcal{N}_{--Y}^{-1} \, 
    \Big( \mathcal{P}^{\mu}_2 \,+\,  \mathcal{P}^{\mu}_5 \Big)\, \bar{v}(p_2) \, \mathbf{\Gamma}_{\mu} \, u(p_1) \, , \nonumber\\
    \mathcal{M}^{[--L]} &= \mathcal{N}_{--L}^{-1} \, 
    \Big( \mathcal{P}^{\mu}_3 \,+\,  \mathcal{P}^{\mu}_6 \Big)\, \bar{v}(p_2) \, \mathbf{\Gamma}_{\mu} \, u(p_1) \, ,
\end{align}
where the superscript, e.g. ++$T$, denotes the respective polarisations of the $b$ quark, anti-$b$ quark and the $Z$ boson. Note that, each polarised amplitude can be decomposed into its vector and axial part,
\begin{align}
\label{eq:pol-decom-vec-axial}
    {\cal M}^{[j]}={\cal M}^{[j]}_{vec}+{\cal M}^{[j]}_{axi} \, ,
\end{align}
where the superscript $[j]$ runs over all the six polarisation configurations shown  in~\eqref{eq:HLamps}. The modulus squares of the normalisation factors involved are  
\begin{align} 
\label{eq:HLampNSQ}
    \mathcal{N}_{++T}^{\, 2} &=  -s  \mathcal{N}_{T}^{\, 2} \, , \nonumber\\
    \mathcal{N}_{++Y}^{\, 2} &= -s \mathcal{N}_{Y}^{\, 2} \, , \nonumber\\
        \mathcal{N}_{++L}^{\, 2} &= -s \mathcal{N}_{L}^{\, 2} \, ,
    %
\end{align}
while $\mathcal{N}_{--T}^{\, 2} = \mathcal{N}_{++T}^{\, 2} \, ,\, 
\mathcal{N}_{--Y}^{\, 2} = \mathcal{N}_{++Y}^{\, 2} \, ,\, 
\mathcal{N}_{--L}^{\, 2} = \mathcal{N}_{++L}^{\, 2}$. The ${\cal N}^2_{T,Y,L}$ are defined in \eqref{eq:ZpolvecNSQ}. 
The complex phase factor of a helicity amplitude is unphysical by itself, and they can be freely set among those associated with independent external polarisation states. 
For definiteness, we choose the positive real-valued roots of the expressions in \eqref{eq:HLampNSQ} when evaluating \eqref{eq:HLamps} in the physical regions.

The scattering amplitude in \eqref{eq:bbZHamplitude} can be written in terms of the polarised amplitudes in \eqref{eq:HLamps} (albeit, with a bit of abuse of notation) as follows:
\begin{align}
\label{eq:reln-mat-polMat}
    |{\cal M}\rangle = \sum_{j=1}^6 e^{i \phi_j} |{\cal M}^{[j]} \rangle \, ,
\end{align}
where $[j]$ runs over all the six polarised states listed in \eqref{eq:HLamps} and $\phi_j$ are the aforementioned corresponding unphysical phase factors. In computing any physical quantity where the 
 squared modulus of the matrix element enters, this phase factor does not contribute owing to the orthogonality of the six polarised amplitudes:
\begin{align}
\label{eq:sqM}
    \Big|{\cal M}\Big|^2=\sum_{j=1}^6 \left|e^{i \phi_j} |{\cal M}^{[j]}\rangle\right|^2=\sum_{j=1}^6 \left|{\cal M}^{[j]}\right|^2\,.
\end{align}

Regarding the polarisation states of the (massive) $Z$ boson, its helicity states 
can be constructed optionally, as circular polarisation states from the linear ones, 
e.g. 
\begin{align}
  \begin{split}
    \varepsilon^\mu_{\pm}(q_1) &= \frac{1}{\sqrt{2}} \left( \varepsilon_T^\mu \pm i \varepsilon_Y^\mu \right).
  \end{split}
\end{align}

\subsection{Projectors for Conventional Form Factors}
\label{sec:projectors_FF}

For the purpose of cross checking the results obtained with the projectors  \eqref{eq:LPprojectors_canonical} as prescribed in the preceding section, 
we performed also the conventional form factor decomposition of \eqref{eq:bbZHamplitude}. The following two subsections are thus devoted to a description of 
Lorentz structures included in the decomposition and the derivation of corresponding form factor projectors. 
Special attention is given to discussing a subtlety in the form factor decomposition of \eqref{eq:bbZHamplitude} with axial current vertices in D dimensions.

\subsubsection{The vector part}
\label{sec:projectors_FF_vec}

By Lorentz covariance the amplitude \eqref{eq:bbZHamplitude} can be expressed as a linear combination of a finite Lorentz structure basis at any finite order. These structures are constrained by physical requirements such as on-shell kinematics and symmetries of the dynamics. Here because of the attachment of a Yukawa interaction on the massless $b$ quark line, the power of elementary Dirac matrices in $\mathrm{\mathbf{\Gamma}}^{\mu}$ should be even to have a non-vanishing matrix element between $\bar{v}(p_2)$ and $u(p_1)$ with opposite chirality.
Under the condition of even powers in elementary Dirac matrices, combined with the P-even constraint from the vector coupling, we see that there are only 4 possible linearly independent Lorentz structures. Thus one can write down for the vector part of the amplitude \eqref{eq:bbZHamplitude}:
\begin{align} 
\label{eq:FFD_vec}
\bar{v}(p_2) \, \mathbf{\Gamma}^{\mu}_{vec} \, u(p_1) &= 
    F_{1,vec} \, \bar{v}(p_2) \, u(p_1) \, p_1^{\mu} 
    \,+\, F_{2,vec} \, \bar{v}(p_2) \, u(p_1) \, p_2^{\mu} \nonumber\\
    &\,+\, F_{3,vec} \, \bar{v}(p_2) \, u(p_1) \, q_1^{\mu}
    \,+\, F_{4,vec} \, \bar{v}(p_2) \,\gamma^{\mu} \slashed{q}_1 u(p_1) \,,
\end{align}
after taking into account also the equations of motion for the on-shell massless spinors $\bar{v}(p_2)$ and $u(p_1)$.  
Notice that linear completeness of the Lorentz structure basis employed in \eqref{eq:FFD_vec} holds in D dimensions without demanding the transversality of the Z boson's physical polarisation states\footnote{We also do not assume Ward identities, which hold for the vector part but not for the axial part.}, because we prefer to use the simple metric tensor $g_{\mu\nu}$ 
instead of the physical polarisation sum rule for index contraction 
in projections. This choice of Lorentz structures also leads to simple form factors clearly seen at the tree level.
The linear decomposition coefficients, $F_{vec,1}, \cdots, F_{vec,4}$, 
are Lorentz invariant functions of external kinematics, which are often called form factors. In addition these functions depend also on the value of dimensionality in dimensional regularisation, and their concrete expressions depend on the perturbative order at which they are computed.

Once given the linear basis as well as the index contraction rule, one can then compute the Gram matrix of this linear basis, and its inverse gives us the projectors with which one obtains the Lorentz-invariant form factors.
The projectors for the pure vector form factors defined in \eqref{eq:FFD_vec}, derived under the aforementioned conditions, are still compact enough to be documented explicitly:
\begin{align} 
\label{eq:FFD_vec_projectors}
    {\mathbb{P}}^{\mu}_{1,vec} &= \bar u(p_1)
    \Big\{
        (-2 + D) t^2  p_1^{\mu} 
        + \Big(2 (-3 + D) m_z^2 s - (-2 + D) t u\Big) p_2^{\mu} + (2 - D) s t q_1^{\mu} \nonumber\\
        &~~~~~~~~~~~~+ s t \slashed{q_1} \gamma^{\mu} 
    \Big\}
        v(p_2) \frac{1}{{\cal K}_{vec}}\,,\nonumber\\
    {\mathbb{P}}^{\mu}_{2,vec} &= \bar u(p_1)
    \Big\{
        \Big(2 (-3 + D) m_z^2 s - (-2 + D) t u\Big) p_1^{\mu}
        + (-2 + D) u^2 p_2^{\mu}
        + (4 - D) s u q_1^{\mu} \nonumber\\
        &~~~~~~~~~~~~- s u \slashed{q_1} \gamma^{\mu}
    \Big\}
        v(p_2) \frac{1}{{\cal K}_{vec}}\,,\nonumber\\
    {\mathbb{P}}^{\mu}_{3,vec} &= \bar u(p_1)
    \Big\{
        - (-2 + D) s t p_1^{\mu}
        - (-4 + D) s u p_2^{\mu} 
        - (2 - D) s^2 q_1^{\mu} 
        - s^2 \slashed{q_1} \gamma^{\mu} 
    \Big\}
    v(p_2)\frac{1}{{\cal K}_{vec}}\,,\nonumber\\
    {\mathbb{P}}^{\mu}_{4,vec} &= \bar u(p_1)
    \Big\{
         s t p_1^{\mu}
        - s u p_2^{\mu} 
        - s^2 q_1^{\mu} 
        + s^2 \slashed{q_1} \gamma^{\mu} 
    \Big\}
    v(p_2)\frac{1}{{\cal K}_{vec}} \, ,
\end{align}
where
\begin{align}
    {\cal K}_{vec} = 2 (-3 + D) s (m_z^2 s - t u)\,.
\end{align}

With the tensor amplitude reconstructed according to \eqref{eq:FFD_vec} with form factors projected using \eqref{eq:FFD_vec_projectors}, we can then compute the polarised amplitudes as defined in \eqref{eq:HLamps} and compare with the direct computation using \eqref{eq:LPprojectors_canonical}, albeit only for contributions from the vector coupling of $Z$ boson.
In fact it is sufficient, and more convenient, to perform the comparison for the projections $\mathcal{P}^{\mu}_i \, \bar{v}(p_2) \, \mathbf{\Gamma}_{\mu} \, u(p_1)$ with $i = 1,\, 2,\, \cdots, 6$.
To this end, one can first calculate matrix elements of each of the four Lorentz basis structures in \eqref{eq:FFD_vec} between the external state vectors as defined in \eqref{eq:MBR_Zpolvec}. 
Namely one projects each of these Lorentz basis structures onto the list of projectors in \eqref{eq:LPprojectors_canonical}, computed using the D dimensional algebra in exactly the same way as how the six projections $\mathcal{P}^{\mu}_i \, \bar{v}(p_2) \, \mathbf{\Gamma}_{\mu} \, u(p_1)$ are computed.
The resulting matrix provides us with the linear transformation needed to combine the vector form factors in \eqref{eq:FFD_vec} to get the projections $\mathcal{P}^{\mu}_i \, \bar{v}(p_2) \, \mathbf{\Gamma}_{\mu} \, u(p_1)$ (keeping only vector coupling contributions).
These matrix elements are purely rational functions in the Mandelstam variables of external kinematics (and also the dimensionality D in general). 
Multiplying these transformation matrix elements with the respective vector form factors accordingly gives us the projections $\mathcal{P}^{\mu}_i \, \bar{v}(p_2) \, \mathbf{\Gamma}_{\mu} \, u(p_1)$, and subsequently polarised amplitudes in \eqref{eq:HLamps}. We will come back to this in section~\ref{sec:irfr}.
In view of this discussion, it is also worthy of mentioning that to alleviate some of the related difficulties within the conventional form factor decomposition approach, it is advocated in ref.~\cite{Peraro:2019cjj} to combine the step of projecting out form factors with the step of composing helicity amplitudes in one go.
~\\

Because the four basis structures in \eqref{eq:FFD_vec} are linearly complete in D dimensions for $\mathcal{M}_{vec}$ in  \eqref{eq:bbZHamplitude} (regardless of the QCD loop order), the tensor amplitude reconstructed according to \eqref{eq:FFD_vec} with form factors projected using \eqref{eq:FFD_vec_projectors} (retaining their full D dependencies) is thus a faithful representation of $\mathcal{M}_{vec}$ in D dimensions.
Owing to this, the aforementioned comparison can be made directly for the un-renormalised un-subtracted bare virtual amplitudes, and the agreement should be exact with full D dependence. 
This is indeed what we saw in our calculation as will be discussed further in section~\ref{sec:irfr}, which serves as a strong check of the computational setup we have established so far.

In contrast, this kind of agreement is no longer observed regarding the axial part of the amplitude \eqref{eq:bbZHamplitude}, i.e.,~$\bar{v}(p_2) \, \mathbf{\Gamma}^{\mu}_{axi} \, u(p_1)$, to be discussed in detail in the following subsection. This is because in this case it is a non-trivial task to exhibit, beyond the tree level amplitude, the full linearly-complete basis for the (conventional) form factor decomposition in D dimensions (with a non-anticommuting $\gamma_5$ prescription).
However, we will show through this work that as long as one is only concerned with physical quantities 
(or quantities that actually contribute to physical observables) it is not necessary to find out and use such an ``ultimate'' basis for form factor decomposition in a calculation done with dimensional regularisation.

\subsubsection{The axial part}
\label{sec:projectors_FF_axi}

Since we are employing a constructive prescription of $\gamma_5$ in DR, i.e.~\eqref{eq:gamma5}, where the full anti-commutativity is sacrificed (in exchange of the cyclicity of trace), algebraic equivalences between expressions in 4 dimensions under the condition of a fully anti-commuting $\gamma_5$ may no longer hold true for their D-dimensional counterparts.
Indeed the manual symmetrisation needed in the proper definition of an Hermitian axial current in DR, as reflected in \eqref{eq:axial-symm}, is already a prominent example of such a point. 
However, full anti-commutativity of $\gamma_5$ is routinely exploited when establishing a linearly independent (and complete) basis of Lorentz structures for amplitudes involving axial vertices in 4 dimensions (in addition to the basic Dirac algebra and also on-shell equations of motion).
This implies that a form factor decomposition basis that is claimed to be linearly independent and complete in 4 dimensions for an amplitude involving axial vertices may not necessarily remain truly linearly complete for this amplitude regularised in D dimensions. As a matter of fact, such an interesting point can be verified straightforwardly for $\mathcal{M}_{axi}$ in  \eqref{eq:bbZHamplitude}  already at the tree level.

Since this subtle point has not yet been discussed much in the literature, let us be more specific about it here.
First of all, following the reasoning similar to what was done for the vector part in the previous subsection, we could write down the following four linearly independent Lorentz structures for the axial part of the amplitude \eqref{eq:bbZHamplitude}
\begin{align} 
\label{eq:FFD_axi_4}
\Big\{
\bar{v}(p_2) \,\gamma_5\, u(p_1) \, p_1^{\mu} \,,\, 
\bar{v}(p_2) \,\gamma_5\, u(p_1) \, p_2^{\mu} \,,\, 
\bar{v}(p_2) \,\gamma_5\, u(p_1) \, q_1^{\mu} \,,\, 
\bar{v}(p_2) \,\gamma^{\mu} \gamma_5\, \slashed{q}_1 u(p_1)
\Big\} \,,
\end{align}
which is linearly complete in 4 dimensions under the condition of even powers in elementary Dirac matrices combined with the P-odd constraint from the axial coupling. One then substitutes the (non-anticommuting) definitions for the bookkeeping $\gamma_5$ in \eqref{eq:FFD_axi_4}, i.e.,~\eqref{eq:gamma5} and \eqref{eq:gamma5-axial}. This yields a list of structures to be used in decomposing $\mathcal{M}_{axi}$.
Once given the linear basis \eqref{eq:FFD_axi_4}, one can then compute the Gram matrix of this linear basis whose inverse gives us the projectors with which one obtains the corresponding Lorentz-invariant form factors, similar to the vector part. In particular, pairs of 
Levi-Civita tensors will be contracted in accordance with \eqref{eq:epsilon-contract} with the resulting space-time metric tensors assumed to be D dimensional.
The projectors for the form factors corresponding to axial Lorentz structures with their bookkeeping form given in \eqref{eq:FFD_axi_4}, derived under the aforementioned conditions, read as:
\begin{align}
\label{eq:FFD_axial_projectors}
    {\mathbb{P}}^{\mu}_{1,axi} &= - {\bar u}(p_1)  {\mathbold \epsilon }_{\gamma\gamma\gamma\gamma} v(p_2) \Bigg\{
    (-1 + D)^2 (48 - 14 D + D^2) (m_z^2 - u)^2 p_1^{\mu} \nonumber\\
    &~~~~~~~~~~+ \Big( (176 + 22 D - 69 D^2 + 16 D^3 - D^4) m_z^4 \nonumber\\ 
    &~~~~~~~~~~+ 2 (-148 - 54 D + 73 D^2 - 16 D^3 + D^4) m_z^2 s \nonumber\\
    &~~~~~~~~~~+ (176 + 22 D - 69 D^2 + 16 D^3 - D^4) t u \nonumber\\
    &~~~~~~~~~~- (176 + 22 D - 69 D^2 + 16 D^3 - D^4) m_z^2 (t + u)\Big) p_2^{\mu} \nonumber\\
    &~~~~~~~~~~- (-1 + D)^2 (48 - 14 D + D^2) s (m_z^2 - u) q_1^{\mu} \Bigg\} \frac{12}{  (8 - 9 D + D^2) s^2 {\cal K}_{axi}} \nonumber\\
    &+ {\bar u}(p_1) {\slashed{q_1}}  {\mathbold \epsilon }^{\gamma\gamma\gamma\mu} v(p_2)
    \Bigg\{ (-56 + 71 D - 16 D^2 + D^3) s (m_z^2 - u) \Bigg\}\frac{12}{  (8 - 9 D + D^2) s^2 {\cal K}_{axi}}\,,\nonumber\\
    {\mathbb{P}}^{\mu}_{2,axi} &=  {\bar u}(p_1)  {\mathbold \epsilon }_{\gamma\gamma\gamma\gamma} v(p_2) \Bigg\{
    \Big((-176 - 22 D + 69 D^2 - 16 D^3 + D^4) m_z^4 \nonumber\\
    &~~~~~~~~~~-- 2 (-148 - 54 D + 73 D^2 - 16 D^3 + D^4) m_z^2 s \nonumber\\
    &~~~~~~~~~~+ (-176 - 22 D + 69 D^2 - 16 D^3 + D^4) t u \nonumber\\
    &~~~~~~~~~~- (-176 - 22 D + 69 D^2 - 16 D^3 + D^4) m_z^2 (t + u)\Big) p_1^{\mu} \nonumber\\ 
    &~~~~~~~~~~- (m_z^2 - t)^2 (-144 - 62 D + 77 D^2 - 16 D^3 + D^4) p_2^{\mu} \nonumber\\
    &~~~~~~~~~~- (m_z^2 - t)(120 + 86 D - 77 D^2 + 16 D^3 - D^4) s q_1^{\mu}
    \Bigg\} \frac{12}{  (8 - 9 D + D^2) s^2 {\cal K}_{axi}}\nonumber\\
    &+ {\bar u}(p_1) {\slashed{q_1}}  {\mathbold \epsilon }^{\gamma\gamma\gamma\mu} v(p_2) \Bigg\{
     (-1 + D) (m_z^2 - t) 
    \Bigg\} \frac{12}{s {\cal K}_{axi}}\,,\nonumber\\
    {\mathbb{P}}^{\mu}_{3,axi} &=- {\bar u}(p_1)  {\mathbold \epsilon }_{\gamma\gamma\gamma\gamma} v(p_2) \Bigg\{
    -(-1 + D)^2 (48 - 14 D + D^2) (m_z^2 - u) p_1^{\mu} \nonumber\\
    &~~~~~~~~~~- (-120 - 86 D + 77 D^2 - 16 D^3 + D^4) (m_z^2 - t) p_2^{\mu}\nonumber\\
     &~~~~~~~~~~+ (-1 + D)^2 (48 - 14 D + D^2) s q_1^{\mu} \Bigg\} \frac{12}{  (8 - 9 D + D^2) s {\cal K}_{axi}}\nonumber\\
    &- {\bar u}(p_1) {\slashed{q_1}}  {\mathbold \epsilon }^{\gamma\gamma\gamma\mu} v(p_2) \Bigg\{
     (-56 + 71 D - 16 D^2 + D^3) s 
    \Bigg\} \frac{12}{(8 - 9 D + D^2) s {\cal K}_{axi}}\,,\nonumber\\
    {\mathbb{P}}^{\mu}_{4,axi} &= - {\bar u}(p_1)  {\mathbold \epsilon }_{\gamma\gamma\gamma\gamma} v(p_2) \Bigg\{ -(-7 + D) (m_z^2 - u) p_1^{\mu} \nonumber\\
    &~~~~~~~~~~- (-1 + D) (m_z^2 - t) p_2^{\mu} + (-7 + D) s q_1^{\mu}
    \Bigg\}\frac{12}{s {\cal K}_{axi}} \nonumber\\
    &- {\bar u}(p_1) {\slashed{q_1}}  {\mathbold \epsilon }^{\gamma\gamma\gamma\mu} v(p_2) (8 - 9 D + D^2)\frac{3}{ {\cal K}_{axi}} \, ,
\end{align}
where 
\begin{align}
&{\cal K}_{axi}=(-888 + 416 D + 560 D^2 - 515 D^3 + 159 D^4 - 21 D^5 + D^6)\nonumber\\
&~~~~~~~~~~(m_z^4 + t u - m_z^2 (s + t + u))\,,\nonumber\\
&{\mathbold \epsilon}^{\gamma\gamma\gamma\mu} \equiv - \frac{i}{6} \epsilon^{\nu\rho\sigma\mu} \gamma_{\nu}\gamma_{\rho}\gamma_{\sigma}\,.
\end{align}

The Gram matrix associated with \eqref{eq:FFD_axi_4} used in deriving the above projectors has a matrix rank 4, confirming the fact that these four structures \eqref{eq:FFD_axi_4} are linearly independent (in D dimensions).
Now, to address the question of whether $\mathcal{M}_{axi}$ lives in a space linearly spanned by \eqref{eq:FFD_axi_4} in D dimensions, we enlarge this list by appending the tree level expression  $\mathcal{M}_{axi}$ directly given by its Feynman-diagrammatic representation with the axial current regularised as described in section~\ref{sec:setup}.
We then compute the Gram matrix of this enlarged list of five elements in exactly the same way as before, and we find out that the matrix rank is increased to 5 rather than staying at 4. This linear dependence test clearly implies that the tree-level axial amplitude $\mathcal{M}_{axi}$ (in D dimensions) is not linearly dependent on the four structures \eqref{eq:FFD_axi_4} according to the usual D-dimensional computational prescription.
However, the similar kind of linear dependence test can be performed for the vector part as well, and there we do see that the matrix rank 
of the accordingly enlarged list is not increased but stays at 4, 
confirming the fact that the $\mathcal{M}_{vec}$ does live in the space linearly spanned by \eqref{eq:FFD_vec} in a D-dimensional computation.
At least for $\mathcal{M}_{axi}$ at the tree level, it is not too difficult to find out a linearly complete Lorentz structure basis in D dimensions for it 
(e.g.~simply enlarging the basis \eqref{eq:FFD_axi_4} by collecting additional tensor structures until enough such as those appearing in the tree-level amplitude as long as they are linearly independent under the algebra in use). 
This leads to a larger linear decomposition basis, which in turn leads to projectors more complicated than \eqref{eq:FFD_axial_projectors}. 
However, by the argument made above, a priori, a D-dimensional linearly-complete basis at the tree level is again not guaranteed to be linearly complete and faithful for $\mathcal{M}_{axi}$ at loop orders. Such an issue is kind of similar to what was already observed in the D-dimensional form factor decomposition of four-quark scattering amplitude $q \bar{q} \rightarrow Q\bar{Q}$ in~\cite{Glover:2004si,Abreu:2018jgq}, albeit of a different technical origin.

There is also another manifestation of this important point in the case at hand, which one can easily examine. One can first apply the thus-derived projectors 
\eqref{eq:FFD_axial_projectors} onto $\mathcal{M}_{axi}$ at the tree level to get the corresponding axial form factors. 
Then the axial part of the amplitude 
\eqref{eq:bbZHamplitude} can be reconstructed similarly to what was done for the vector part in accordance with the formula \eqref{eq:FFD_vec}.
However, one can verify explicitly that with the amplitude reconstructed in this way one can not recover the unpolarised squared modulus of the axial part of amplitude \eqref{eq:bbZHamplitude} (at the tree level) computed using the physical polarisation sum rule in CDR directly from its original Feynman diagrammatic representation. This tells us again that the computation of the tensor amplitude with \eqref{eq:FFD_axi_4} is not algebraically identical to the original $\mathcal{M}_{axi}$ (even at the tree level) in D dimensions. Because of this we can not write down an exact decomposition identity for $\mathcal{M}_{axi}$ just in terms of the basis \eqref{eq:FFD_axi_4} in analogy to its vector counterpart \eqref{eq:FFD_vec}.~\\

Moreover, the projectors in \eqref{eq:FFD_axial_projectors} are not very convenient to use in practice, especially if one considers some more complicated versions with additional D-dimensional linearly independent Lorentz structures incorporated in the form factor decomposition basis. 
However, a priori, it is not clear whether omitting additional (evanescent) Lorentz structures in the form factor decomposition of a D-dimensional amplitude, i.e.,~using a D-dimensional linearly-incomplete decomposition basis like \eqref{eq:FFD_axi_4} for $\bar{v}(p_2) \, \mathbf{\Gamma}^{\mu}_{axi} \, u(p_1)$ at loop orders, would still always lead to correct results,
because for sure the so reconstructed amplitude is not algebraically identical to the original defining form (given by Feynman diagrams). Bold moves of this kind have already been tried out, e.g.~in the computation of $q\bar{q} \rightarrow t\bar{t}$ amplitudes to two-loop order in ref.~\cite{Chen:2017jvi}.
On top of this undecided situation, there is also no clear statement in the literature about whether one can always set the dimension variable D = 4, i.e. in the curly brackets of the expressions in \eqref{eq:FFD_axial_projectors}, and still expect with confidence that the amplitude reconstructed in this way yields correct results.

Let us now summarize the issues we have just discussed regarding the form factor decomposition of an amplitude involving axial vertices in D dimensions,
in particular the expression $\bar{v}(p_2) \, \mathbf{\Gamma}^{\mu}_{axi} \, u(p_1)$ at hand.
\begin{enumerate}
\item 
It is not easy to construct the full D-dimensional linearly complete basis for the form factor decomposition of $\bar{v}(p_2) \, \mathbf{\Gamma}^{\mu}_{axi} \, u(p_1)$ at loop orders.
\item 
Even with just an incomplete basis, keeping the full D dependence in the derivation of corresponding projectors could lead to expressions with too complicated D dependence to use in practice, e.g.~\eqref{eq:FFD_axial_projectors}. 
\item 
Setting D=4 within the curly brackets of the form factor projectors \eqref{eq:FFD_axial_projectors} could simplify the expressions a bit. However, it is not known whether this is legitimate in general for the form factor decomposition. 
\end{enumerate}
Concerns like these were among the motives that drove us to take the approach as prescribed in ref.~\cite{Chen:2019wyb}.
However, it is still very interesting to investigate whether the aforementioned bold moves, especially combining the second and third point above, would work all the way through. 
This is then what we did for the axial part of amplitude \eqref{eq:bbZHamplitude}, in addition to the computation using the projectors discussed in section~\ref{sec:projectors_LP}, in order to answer this question. Namely, not only did we choose a set of form factor decomposition structures that are known to be linearly incomplete in D dimensions (albeit, linearly independent and complete in 4 dimensions), but we also set manually all explicit D in the resulting form factor projectors to be 4.
In other words, for the form factor decomposition of $\mathcal{M}_{axi}$ we used \eqref{eq:FFD_axial_projectors} with D=4 in the curly brackets, i.e.,~a set of form factor projectors that are essentially identical to the respective expressions that would be used in a four-dimensional form factor decomposition\footnote{Let us stress that in this acrobatic axial form factor decomposition there is no imposition of any explicit dimensional splitting, which would have allowed one to keep strictly only 4 dimensional d.o.f. in the external projectors (clearly separated from all orthogonal evanescent ones) such as typically done in the HV scheme.}. Namely we \textit{define} our ``axial form factors'' (throughout this article) as 
\begin{align} 
\label{eq:FFD_axi_def}
F_{1,axi} &\equiv {\mathbb{P}}^{[\text{4}],\,\mu}_{1,axi} \, \bar{v}(p_2) \, \mathbf{\Gamma}^{\nu}_{axi} \, u(p_1)\,g_{\mu \nu}\,\nonumber\\
F_{2,axi} &\equiv {\mathbb{P}}^{[\text{4}],\,\mu}_{2,axi} \, \bar{v}(p_2) \, \mathbf{\Gamma}^{\nu}_{axi} \, u(p_1)\,g_{\mu \nu}\,\nonumber\\
F_{3,axi} &\equiv {\mathbb{P}}^{[\text{4}],\,\mu}_{3,axi} \, \bar{v}(p_2) \, \mathbf{\Gamma}^{\nu}_{axi} \, u(p_1)\,g_{\mu \nu}\,\nonumber\\
F_{4,axi} &\equiv {\mathbb{P}}^{[\text{4}],\,\mu}_{4,axi} \, \bar{v}(p_2) \, \mathbf{\Gamma}^{\nu}_{axi} \, u(p_1)\,g_{\mu \nu}\,
\end{align}
where the $[\text{4}]$ in the superscript denotes the fact that we set  D = 4 in the original ${\mathbb{P}}^{\mu}_{i,axi}$. Note that, however, when applying the so determined ``axial form factor projectors'' to $\mathcal{M}_{axi}$, 
any pair of Levi-Civita symbols will always be contracted in accordance with  \eqref{eq:epsilon-contract} where all indices of the resulting metric tensors are  D-dimensional~\cite{Zijlstra:1992kj}. Subsequently we build up an intermediate axial amplitude $\tilde{\mathcal{M}}_{axi}^{\mu}$ defined as 
\begin{align} 
\label{eq:FFD_axi_amp}
\tilde{\mathcal{M}}_{axi}^{\mu} &\equiv 
    F_{1,axi} \, \bar{v}(p_2) \,\gamma_5\, u(p_1) \, p_1^{\mu} 
    \,+\, F_{2,axi} \, \bar{v}(p_2) \,\gamma_5\, u(p_1) \, p_2^{\mu} \nonumber\\
    &\,+\, F_{3,axi} \, \bar{v}(p_2) \,\gamma_5\, u(p_1) \, q_1^{\mu}
    \,+\, F_{4,axi} \, \bar{v}(p_2) \,\gamma^{\mu} \, \gamma_5\, \slashed{q}_1 u(p_1) \,.
\end{align}
This quantity $\tilde{\mathcal{M}}_{axi}^{\mu}$ is not
algebraically identical (or faithful) to the original Feynman amplitude $\mathcal{M}_{axi}^{\mu} = \bar{v}(p_2) \, \mathbf{\Gamma}^{\mu}_{axi} \, u(p_1)$ in D dimensions, due to issues discussed above, while we expect that the two should eventually lead to the same properly defined finite remainder in four dimensions.

Indeed, with well established results for polarised amplitudes obtained using physical projectors defined in section~\ref{sec:projectors_LP}, not relying on any explicit Lorentz tensor decomposition of the original Feynman amplitude (both conceptually and technically), we have verified that eventually the same finite remainders for  $\mathcal{M}_{axi}$ were obtained with both approaches,
and hence we confirm a positive answer to the question raised above for the process \eqref{eq:process}. We will come back to this later in section~\ref{sec:irfr}. The comparison regarding these axial amplitudes can proceed as described at the end of the last subsection, but with one crucial difference compared to their vector counterparts: the agreement can only be expected at the level of properly defined finite remainders in four dimensions, due to the subtle points explained above.

\section{UV Renormalisation}
\label{sec:uv}

The bare scattering amplitude of the process \eqref{eq:process} beyond the leading order (LO) is not finite, and it contains poles in dimensional regulator $\epsilon (\equiv 1/2(4-D)$) arising from ultraviolet, soft and collinear regions of the loop momenta. 
The UV renormalisation of the amplitude \eqref{eq:bbZHamplitude} is done in the modified minimal subtraction ($\overline{\mathrm{MS}}$) scheme.
To be more specific, the UV divergences in the vector part of the amplitude \eqref{eq:bbZHamplitude} in massless QCD are handled by the QCD coupling constant renormalisation through
\begin{align}
\label{eq:qcd-renorm}
    {\hat a}_s S_{\epsilon}=a_s(\mu_R^2) Z_{a_s}(\mu_R^2) \left(\frac{\mu^2}{\mu_R^2}\right)^{-\epsilon}
\end{align}
and the renormalisation of the Yukawa coupling through 
\begin{align}
\label{eq:yukawa-renorm}
    {\hat \lambda}_{b} S_{\epsilon}=\lambda_b(\mu_R^2) Z_{\lambda}(\mu_R^2) \left(\frac{\mu^2}{\mu_R^2}\right)^{-\epsilon} \, ,
\end{align}
where $S_{\epsilon} \equiv \exp\left[\epsilon(\ln 4\pi-\gamma_E)\right]$ with Euler constant $\gamma_E=0.5772...$. In \eqref{eq:qcd-renorm}, \eqref{eq:yukawa-renorm} and throughout this article, hat ($\;\hat{}\;$) represents the bare quantity, $a_s \equiv \alpha_s/4\pi$ is the strong coupling constant, $\mu$ is an auxiliary mass-dimensionful parameter introduced in dimensional regularisation to keep the coupling constants dimensionless and $\mu_R$ is the usual renormalisation scale. 
These two QCD renormalisation constants are given  to two-loop order in the  $\overline{\rm MS}$
scheme by~\cite{Tarasov:1980au}: 
\begin{align}
\label{eq:ren-const}
    &Z_{a_s}(\mu_R^2)=1+a_s(\mu_R^2) \Bigg(  \frac{1}{\epsilon}\left\{ -\frac{11}{3} C_A+\frac{2}{3}n_f\right\}\Bigg)+a_s^2(\mu_R^2)
    \Bigg( \frac{1}{\epsilon^2} \Bigg\{ \frac{121}{9}C_A^2-\frac{44}{9}C_A n_f+\frac{4}{9}n_f^2 \Bigg\}\nonumber\\
    &\qquad\qquad\qquad\qquad\qquad~~~~~~~~~~~~~+ \frac{1}{\epsilon} \Bigg\{-\frac{17}{3} C_A^2+\frac{5}{3} C_A n_f + C_F n_f \Bigg\} \Bigg)\,,\nonumber\\
    &Z_{\lambda}(\mu_R^2)=1+a_s(\mu_R^2)\Bigg(-\frac{3}{\epsilon^2} C_F\Bigg)+a_s^2(\mu_R^2) \Bigg(\frac{1}{\epsilon^2} \left\{ \frac{11}{2} C_A C_F - C_F n_f  + \frac{9}{2} C_F^2 \right\} \nonumber\\
    &\qquad\qquad\qquad\qquad\qquad~~~~~~~~~~~~+ \frac{1}{\epsilon}  \left\{  \frac{5}{6} C_F n_f  -\frac{97}{12} C_A C_F - \frac{3}{4} C_F^2 \right\} \Bigg) \,.
\end{align}
%
%
\begin{align}
\label{eq:beta}
    \beta_0=\frac{11}{3} C_A -\frac{2}{3} n_f\,.
\end{align}
The quantities $C_A=N_c$ and $C_F=(N_c^2-1)/(2 N_c)$ are the eigenvalues of the quadratic Casimir operators in the adjoint and the fundamental representations of the SU($N_c$) gauge group, respectively.

However, \eqref{eq:qcd-renorm} and \eqref{eq:yukawa-renorm} are not sufficient to properly UV renormalise the axial part of \eqref{eq:bbZHamplitude}  
with $\gamma_5$ regularised in DR as described in section~\ref{sec:setup}.  
The particular prescription for the $\gamma_5$-related quantities,
such as \eqref{eq:gamma5-axial} for the axial current which violates the anti-commutativity of $\gamma_5$ with $\gamma^{\mu}$, gives rise to some additional spurious UV poles in $\epsilon$ in dimensional regularisation which has to be removed manually~\cite{Chanowitz:1979zu,Trueman:1979en}. 
Manifestation of this issue shows up as the loss of correct chiral Ward identities for axial currents in the original prescriptions by 't Hooft-Veltman~\cite{tHooft:1972tcz} and Breitenlohner-Maison~\cite{Breitenlohner:1977hr} (at higher loop orders).
This amendment is typically realized in form of some additional UV renormalisation constants~\cite{Trueman:1979en,Kodaira:1979pa,Larin:1991tj,Larin:1993tq} introduced specifically for the axial part of the amplitude (on top of the aforementioned renormalisations common to the vector part). 
The concrete expressions of these additional finite renormalisation constants depend on the treatment of 
the Levi-Civita tensor appearing in the constructive expression eq.(\ref{eq:gamma5}) for the non-anticommutating $\gamma_5$. 
In practice, it is convenient to have this rectification done with the so-called flavour singlet and non-singlet Feynman diagrams separated, which we now turn to in the following subsections.

\subsection{UV Renormalisation of the Flavour Non-Singlet Current}
\label{ss:UV-ns}

\begin{figure}[h!]
\begin{center}
\includegraphics[scale=0.55]{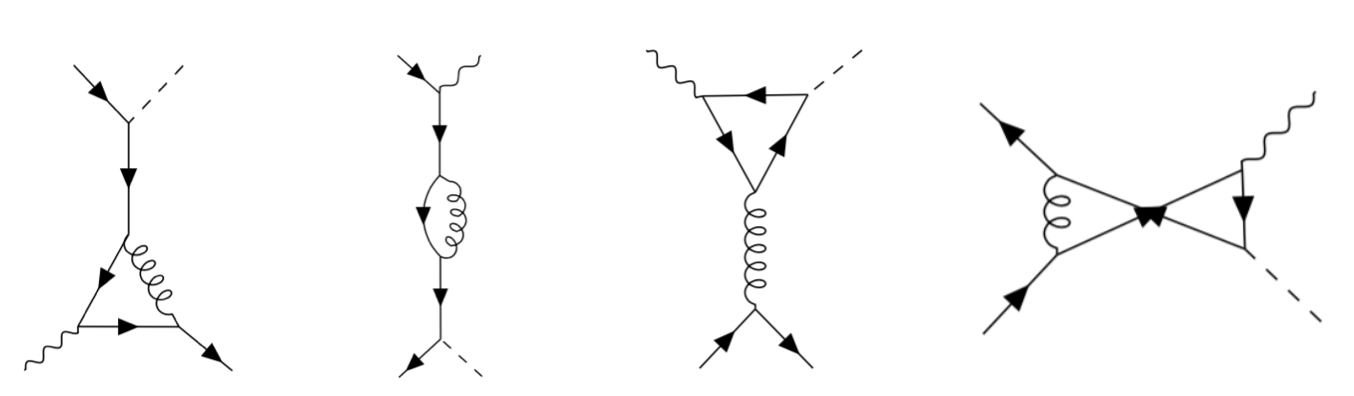}
\caption{Examples of Feynman diagrams at one-loop order. The open solid lines in these diagrams represent the $b$ quark with an arrow indicating the flow of its charge. The open dashed (wavy) line denotes the external Higgs ($Z$) boson, while gluons are represented by the curly lines.}
\label{dia:oneloop}
\end{center}
\end{figure}

\begin{figure}[htb]
\begin{center}
\includegraphics[scale=0.55]{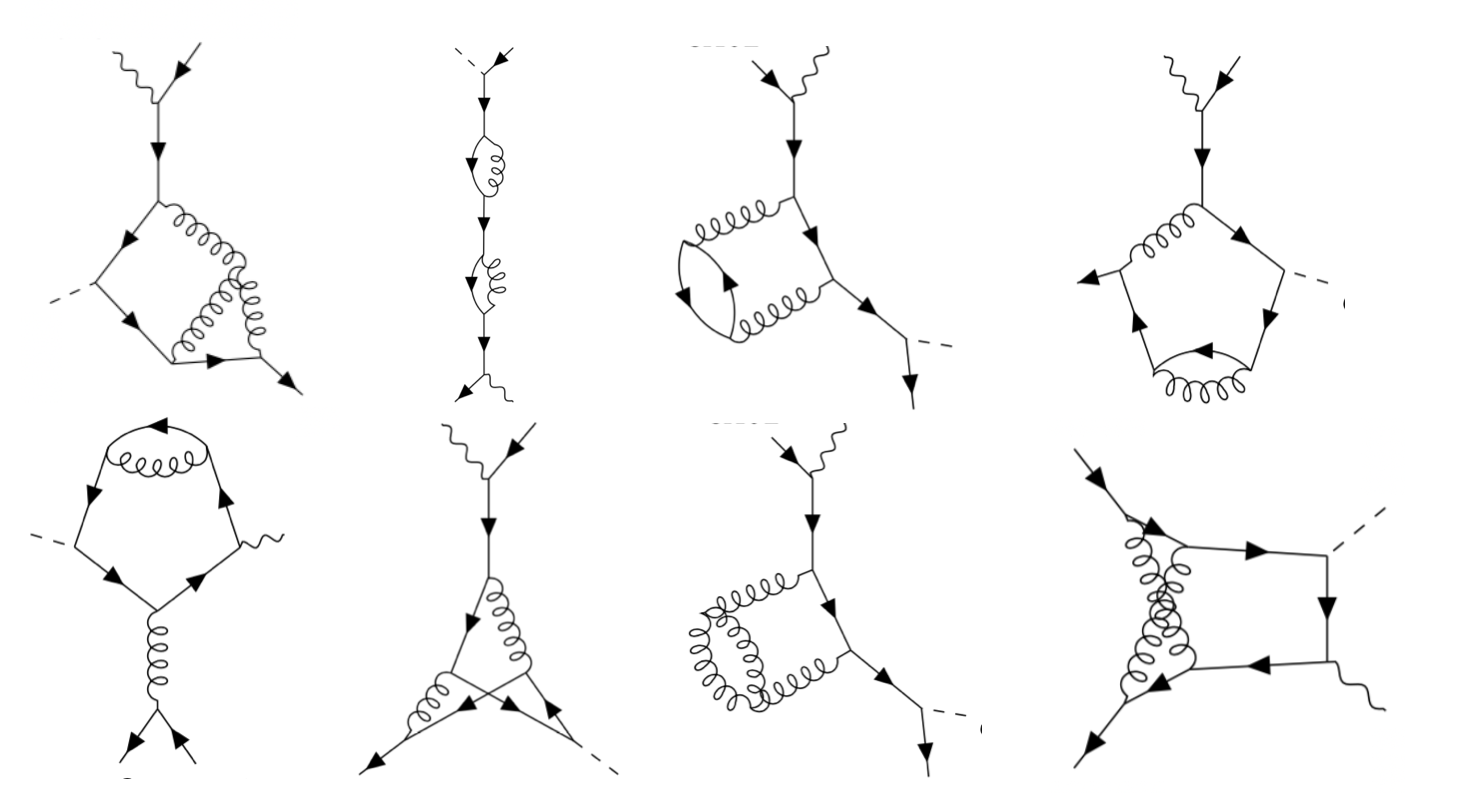}
\caption{Examples of non-singlet Feynman diagrams at two-loop order.}
\label{dia:non-singlet}
\end{center}
\end{figure}

Samples of the flavour non-singlet, or non-anomalous, set of Feynman diagrams are shown in figure~\ref{dia:oneloop} at one-loop order and in figure~\ref{dia:non-singlet} at two-loop order, where there is no appearance of the closed triangle fermion loop with axial $Z$ coupling vertex (related to the quantum anomaly to be addressed in the next subsection). 
For these diagrams, the axial part of the coupling between the $Z$ boson and the $b$ quark takes place through the flavour non-singlet axial current
\begin{align}
\label{eq:ns-axial-current}
    J^{ns}_{\mu,A}(x) = \bar\psi(x) \gamma_{\mu}\gamma_5  I_{3} \psi(x)=\frac{i}{6} \epsilon_{\mu\nu\rho\sigma} \bar\psi(x) \gamma^{\nu}\gamma^{\rho}\gamma^{\sigma} I_3 \psi(x) \, ,
\end{align}
where $I_3$ denotes the third component of the (weak) isospin operator. 
The appearance of additional spurious UV divergences as a result of 
using the non-anticommutating prescription of $\gamma_5$ in dimensional regularisation needs to be compensated by performing an overall $\overline{\mathrm{MS}}$ renormalisation through $Z^{ns}_A$ 
and on top of it by multiplying an additional finite renormalisation constant $Z^{ns}_{5,A}$~\cite{Trueman:1979en}:
\begin{align}
\label{eq:renorm-ns-current}
    J^{ns}_{\mu,A}(x) = Z^{ns}_{5,A} Z^{ns}_A {\hat J}^{ns}_{\mu,A}(x)\,,
\end{align}
where, as already mentioned, the hat ( ${\hat{}}$ ) represents the bare operator. 
The aforementioned renormalisation constants up to two-loop order are given by~\cite{Larin:1991tj}: 
\begin{align}
\label{eq:renorm-const-ns}
    &Z^{ns}_A=1+a_s^2(\mu_R^2) \frac{1}{\epsilon} \left( \frac{22}{3} C_F C_A -\frac{4}{3} C_F n_f\right)\,, \nonumber\\
    &Z^{ns}_{5,A}=1+a_s(\mu_R^2) \left( -4 C_F\right)+a_s^2(\mu_R^2) \left( 22 C_F^2-\frac{107}{9} C_F C_A + \frac{2}{9} C_F n_f \right)\,.
\end{align}
The properly UV renormalised non-singlet axial current in \eqref{eq:renorm-ns-current} is non-anomalous and 
should be conserved for massless quarks, i.e.~the standard Ward identity holds and the renormalisation invariance of the current is restored (just like its vector counterpart).
While, due to the presence of the non-zero Yukawa coupling $\lambda_b$
in the amplitude \eqref{eq:bbZHamplitude} under consideration, the current $\bar{v}(p_2) \, \mathbf{\Gamma}^{\mu}_{axi} \, u(p_1)$ 
(which is a Green correlation function involving the axial current operator $J^{ns}_{\mu,A}\,$) is no longer conserved even though the $b$ quark mass 
is set to be zero.  
This can be verified explicitly by replacing the polarisation vector of the $Z$  boson by its momentum in the amplitude $\mathcal{M}_{axi}$ which yields $q_{1,\mu} \, \bar{v}(p_2) \, \mathbf{\Gamma}^{\mu}_{axi} \, u(p_1) \neq 0$, and this expression is proportional to $\lambda_b$. This non-conservation is ``classically expected" and hence not regarded as a ``quantum anomaly", which we will turn to in the next subsection. While on the other hand $q_{1,\mu} \, \bar{v}(p_2) \, \mathbf{\Gamma}^{\mu}_{vec} \, u(p_1) = 0$ is still true.

\subsection{UV Renormalisation of the Flavour Singlet Current and ABJ Anomaly}
\label{ss:UV-s}

\begin{figure}[htbp]
\begin{center}
\includegraphics[scale=0.5]{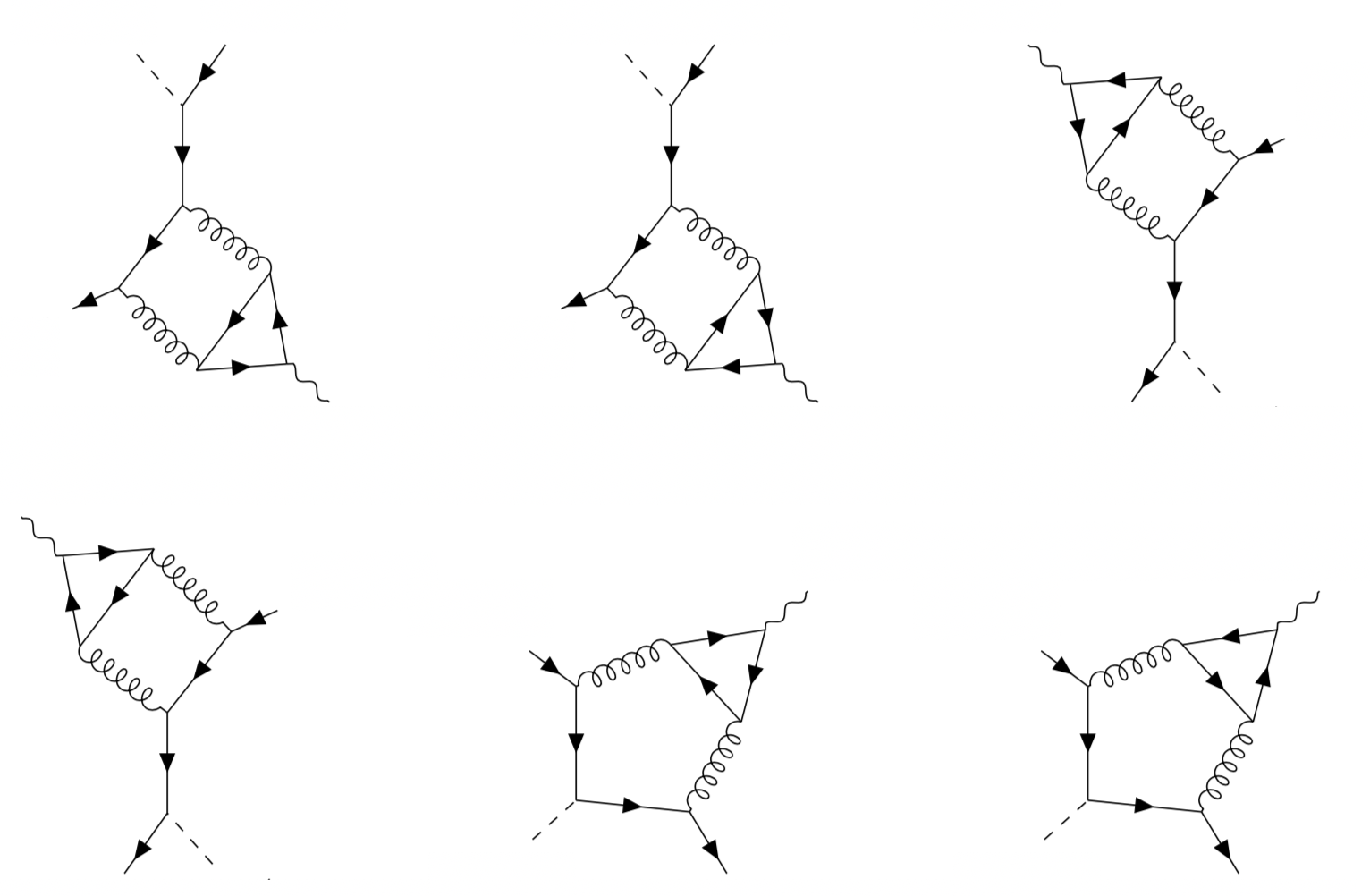}
\caption{Singlet Feynman diagrams at two-loop order.}
\label{dia:singlet}
\end{center}
\end{figure}

The flavour-singlet (or anomalous) Feynman diagrams, characterised by featuring a triangle fermion loop with axial $Z$ coupling, start to appear at two-loop order for the process in question. They are shown in figure~\ref{dia:singlet}.
In this class of diagrams the $Z$ boson couples to the quark circulating around the triangle fermion loop.
The quark flavour running through the triangle fermion loop can be any of the $n_f$ flavours, as it is disconnected from the external open $b$ quark line. The $Z$ boson's vector coupling with the quarks does not contribute in this set of diagrams due to the Furry's theorem, and thus only the axial part survives.
Moreover, since the weak-isospin quantum number of up (charm) and down (strange) quarks differ just by a sign, the contributions coming from the first and second generations of the quarks vanish in this set of diagrams. 
The only non-zero contribution to this class of anomalous diagrams in $n_f = 5$ massless QCD comes from the bottom quark triangle loop\footnote{In the full QCD with the (massive) top quark, there are still non-vanishing contributions from this class of diagrams as the third generation of quarks is not degenerate in mass.}(in absence of the top quark). 
The interaction between the bottom quark of this triangle loop and $Z$ boson proceeds via the flavour singlet axial current
\begin{align}
\label{eq:axial-current}
    J^{s}_{\mu,A}(x) = \bar\psi(x) \gamma_{\mu}\gamma_5  \psi(x)=\frac{i}{6} \epsilon_{\mu\nu\rho\sigma} \bar\psi(x) \gamma^{\nu}\gamma^{\rho}\gamma^{\sigma} \psi(x)\,.
\end{align}
To UV renormalise the singlet axial current properly, as in the previous case, in addition to performing an overall $\overline{\mathrm{MS}}$ renormalisation, we also need to multiply with an additional finite renormalisation constant~\cite{Kodaira:1979pa,Collins:1984xc,Larin:1993tq}:
\begin{align}
\label{eq:renorm-s-current}
    J^{s}_{\mu,A}(x) = Z^{s}_{5,A} Z^{s}_A {\hat J}^{s}_{\mu,A}(x)\,.
\end{align}
The renormalisation constants up to two-loop order read as~\cite{Larin:1993tq}:
\begin{align}
\label{eq:ren-constant-s}
	&Z^{s}_{A}=1+a_s^2(\mu_R^2) \frac{3}{\epsilon}  C_F \nonumber\,,\\
	&Z^{s}_{5,A}=1+a_s^2(\mu_R^2) \frac{3}{2}  C_F\,.
\end{align}

Unlike the flavour non-singlet axial current, the singlet axial current does not satisfy the standard Ward identity even in the massless quark limit. This current exhibits anomalous properties which are known as axial or Adler-Bell-Jackiw (ABJ) anomaly~\cite{Adler:1969gk,Bell:1969ts}. The operator relation for the ABJ anomaly of massless axial current reads~\cite{Adler:1969er}:
\begin{align}
\label{eq:ABJ}
    \left(\partial^{\mu} J^{s}_{\mu,A}\right)_R=a_s\frac{1}{2}\left(G\tilde G\right)_R \, ,
\end{align}
where $G\tilde G \equiv \epsilon_{\mu\nu\rho\sigma}G^a_{\mu\nu}G^a_{\rho\sigma}$ and $G^a_{\mu\nu}$ is the gluonic field strength tensor. The subscript $R$ of the composite local operators on both sides of \eqref{eq:ABJ} represents that these composite local operators need to be properly renormalised so that this operator relation holds.
The operator on the left-hand side is renormalised multiplicatively in the same way as the current \eqref{eq:renorm-s-current} itself, while the renormalisation of the $G\tilde G$ can be found in~\cite{Larin:1993tq,Ahmed:2015qpa} in detail. The concrete expression of $Z^{s}_{5,A}$ in \eqref{eq:ren-constant-s} is  determined precisely such that the properly UV renormalised singlet axial current in \eqref{eq:renorm-s-current} has an anomaly that does obey \eqref{eq:ABJ}, i.e.~, that preserves the \textit{one-loop} character of this relation.

\begin{figure}[htbp]
\begin{center}
\includegraphics[scale=0.7]{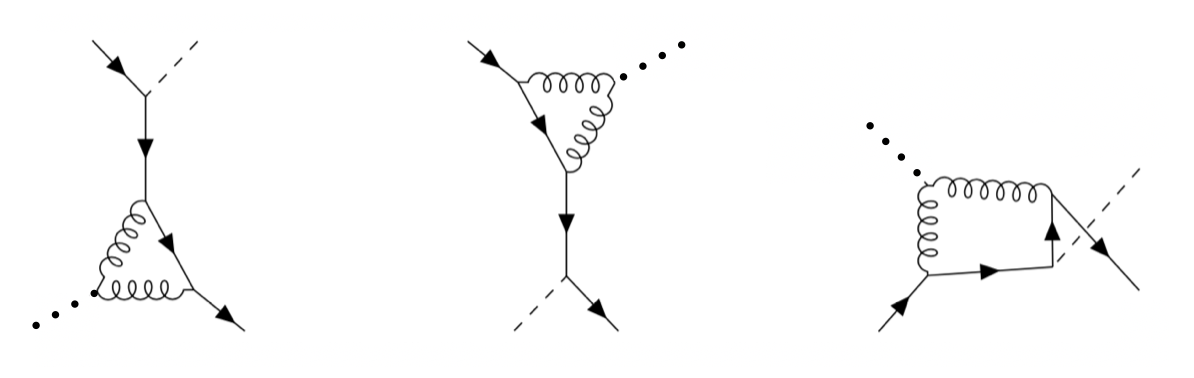}
\caption{Diagrams containing contributions from $G\tilde G$ in \eqref{eq:ABJ}. The dotted line represents a pseudo-scalar.}
\label{dia:bb2ah}
\end{center}
\end{figure}

Technically speaking, \eqref{eq:ABJ} is irrelevant regarding 
the contributions to $\mathcal{M}_{axi}$ from this flavour singlet set of 6 Feynman diagrams all depicted in figure~\ref{dia:singlet}. This is because the polarisation states of the external on-shell $Z$ boson in the physical amplitude from these diagrams, either transversal or longitudinal as listed in \eqref{eq:MBR_Zpolvec}, are always orthogonal to its own momentum. However, we verified this operator level relation explicitly
within our computational setup for the amplitude $\mathcal{M}_{axi}$ 
in order to perform an indirect check of our treatment of these anomalous diagrams.
To be a bit more specific about this, let us denote the contribution to  $\mathcal{M}_{axi}$ from these 6 anomalous Feynman diagrams by $\bar{v}(p_2) \, \mathbf{\Gamma}^{\mu}_{\text{ABJ}} \, u(p_1)\, \varepsilon_{\mu}^{*}$.
The Green correlation function of the external fields with the left-hand side 
of \eqref{eq:ABJ} (or rather, the matrix element of the \textit{external} operator $\left(\partial^{\mu} J^{s}_{\mu,A}\right)_R $ between the vacuum 
and the external on-shell states of our process \eqref{eq:process} excluding the $Z$ boson) reads as $\bar{v}(p_2) \, \mathbf{\Gamma}^{\mu}_{\text{ABJ}} \, u(p_1)\, q_{1,\mu}$. Despite the non-vanishing anomaly $\left(\partial^{\mu} J^{s}_{\mu,A}\right)_R$ at the operator level,  
$\bar{v}(p_2) \, \mathbf{\Gamma}^{\mu}_{\text{ABJ}} \, u(p_1)\, q_{1,\mu}$ would have completely vanished (since $b$ quark is taken massless) if it were not due to the presence of the $b$ quark Yukawa coupling in our calculation.
Replacing $\left(\partial^{\mu} J^{s}_{\mu,A}\right)_R $ by 
the operator on the right-hand side of \eqref{eq:ABJ} leads to 
3 one-loop diagrams, as depicted in figure~\ref{dia:bb2ah}, which are connected to those in figure~\ref{dia:singlet} by shrinking the fermion triangle loop into an effective vertex between two gluons and one imaginary or auxiliary pseudo-scalar.
These one-loop diagrams would have also been vanishing if there would be no Higgs
radiated from the (massless) $b$ quark, simply due to the tension between the angular momentum conservation and chirality conservation along the massless $b$ quark line (without Yukawa coupling). In the end we computed these one-loop 
diagrams and $\bar{v}(p_2) \, \mathbf{\Gamma}^{\mu}_{\text{ABJ}} \, u(p_1)\, q_{1,\mu}$, both of which are proportional to the Yukawa coupling
$\lambda_b$, and found an exact agreement between the two and hence 
verified \eqref{eq:ABJ} in our computational setup.
~\\

In view of the classification of contributing diagrams as discussed above, 
the scattering amplitude \eqref{eq:bbZHamplitude} as well as 
the polarised ones defined in \eqref{eq:HLamps} are thus conveniently 
separated into pieces with their respective renormalisations 
as follows:
\begin{align}
\label{eq:UV-Amp}
    {\cal M}^{[j]}&= 
    {\cal M}^{[j]}_{vec}(a_s(\mu_R^2)) + {\cal M}^{[j]}_{axi}(a_s(\mu_R^2))\nonumber\\
    &= {\cal M}^{[j]}_{vec}(a_s(\mu_R^2)) + {\cal M}^{[j],ns}_{axi}(a_s(\mu_R^2)) 
        + {\cal M}^{[j],s}_{axi}(a_s(\mu_R^2)) \nonumber\\
    &= \hat {\cal M}^{[j]}_{vec}(\hat a_s, \mu^2)
        + Z_5^{ns}(a_s(\mu_R^2)) Z_A^{ns}(a_s(\mu_R^2)) \hat {\cal M}^{[j],ns}_{axi}(\hat a_s,\mu^2)\nonumber\\
        &~~~~~~~~~~~~\qquad\quad+ Z_5^{s}(a_s(\mu_R^2)) Z_A^{s}(a_s(\mu_R^2)) \hat {\cal M}^{[j],s}_{axi}(\hat a_s,\mu^2)\,,
\end{align}
where the superscript $[j]$ runs over all the six polarisation configurations as in~\eqref{eq:HLamps}.
Upon applying the projectors (defined in section \ref{sec:projectors}) onto the Feynman amplitudes, we compute the bare polarised amplitudes $\hat {\cal M}^{[j]}_{vec}(\hat a_s, \mu^2)$, $\hat {\cal M}^{[j],ns}_{axi}(\hat a_s,\mu^2)$ and $\hat {\cal M}^{[j],s}_{axi}(\hat a_s,\mu^2)$. The UV renormalised amplitudes can be expanded in powers of $a_s$ as
\begin{align}
\label{eq:UV-amp-expansion}
    {\cal M}^{[j]}=\lambda(\mu_R^2)\sum_{l=0}^{\infty} a_s^l(\mu_R^2){\cal M}^{[j],(l)}\,.
\end{align}
In this article we will present our analytic results of these polarised amplitudes to two-loop level, i.e.,~${\cal M}^{[j],(0)}$, ${\cal M}^{[j],(1)}$ and ${\cal M}^{[j],(2)}$.

\section{Computation of the Amplitudes}
\label{sec:computation}

Despite the presence of modern techniques based on unitarity (or on-shell cuts), the Feynman diagrammatic approach of computing loop amplitudes still remains a popular choice which is employed in this article, especially in view of many available efficient tools for Feynman diagram generations and manipulations for a large variety of theory models.\footnote{In fact the projection approach employed in this article is also compatible with the (D-dimensional) on-shell cut-based reconstruction method for representing Feynman amplitudes, in the sense that the \textit{symbolic} expressions of Feynman amplitudes in D dimensions used in the projection can be composed via on-shell cut-based reconstruction rather than appealing to their traditional Feynman diagrammatic representations.}. 
The technical aspects of the computation of bare Feynman amplitudes in \eqref{eq:UV-Amp} closely follow the steps used in the recent calculation of doubly massive form factors~\cite{Ahmed:2019upm} in maximally supersymmetric Yang-Mills theory. Feynman diagrams are generated symbolically using \qgraf~\cite{Nogueira:1991ex}. For the non-Drell-Yan type contributions to the process \eqref{eq:process}, there are 2 and 10 diagrams at the tree and one-loop level, respectively. At two-loops, there are 153 flavour non-singlet and 6 flavour singlet diagrams generated in our setup. The symbolically generated diagrams are passed through a series of in-house codes based on \form~\cite{Vermaseren:2000nd} in order to apply the Feynman rules, perform SU($N_c$) colour and D-dimensional Lorentz and Dirac algebras. Upon multiplying the properly constructed  projectors, every projected Feynman amplitude is expressed as a linear combination of a large number of scalar Feynman integrals belonging to the family of four-point amplitudes with two off-shell legs of different virtualities. Using the liberty of transforming the loop momenta, all the scalar Feynman integrals are categorised into three and six different integral families at one- and two-loop order, respectively, with the help of \reduze~\cite{Studerus:2009ye,vonManteuffel:2012np}. These scalar integrals can be reduced to linear combinations of a much smaller number of loop integrals, called master integrals (MI), using integration-by-parts (IBP)~\cite{Tkachov:1981wb,Chetyrkin:1981qh}. To perform the IBP reduction, we use \litered~\cite{Lee:2012cn} along with \mint~\cite{Lee:2013hzt} at one-loop and \kira~\cite{Maierhoefer:2017hyi,Maierhofer:2018gpa} at two-loop level. Upon performing the IBP reduction, we get at the two-loop level 134 MI which are further reduced to a set of 84 independent integrals by taking into account additional relations and crossings of the external momenta.

In a parallel setup, we first obtain the unreduced symbolic form of polarised amplitudes using an extension of the program \gosam~\cite{Cullen:2011ac,Cullen:2014yla,Jones:2016bci} at one-loop and two-loop order. The list of unreduced loop integrals appearing is then extracted and fed to \kira~\cite{Maierhoefer:2017hyi,Maierhofer:2018gpa} to obtain a table of generally usable IBP rules. Insertion of the IBP table and subsequent simplification of rational coefficients in front of MI are performed with an in-house routine based on a parallelised usage of Mathematica and \fermat~\cite{fermat}.
We have checked that the final reduced forms of all projected amplitudes in terms of MI are identical for these two parallel setups.
~\\

The set of independent MI involved in our amplitudes agrees with those in the four-point amplitudes with two massive legs with different virtualities which were first computed in refs.~\cite{Henn:2014lfa,Caola:2014lpa,Papadopoulos:2014hla}. A subset of these MI were computed in refs.~\cite{Chavez:2012kn,Anastasiou:2014nha}. Later, an independent computation of these MI was performed in ref.~\cite{Gehrmann:2015ora} where the solutions are optimised for efficient numerical evaluation. For our present calculation, we use the optimised solutions of the MI computed in ref.~\cite{Gehrmann:2015ora} which are available in \hepforge~\cite{hepforge} in computer readable format. In order to use the optimised solutions, we introduce the same set of dimensionless quantities defined in ref.~\cite{Gehrmann:2015ora} constructed out of the Mandelstam variables \eqref{eq:Mandelstam}
\begin{align}
\label{eq:varch}
    s=m^2(1+x)(1+xy)\,,~~~ t=-m^2xz\,,~~~ q_1^2=m^2\,,~~~ q_2^2=m^2x^2y\,.
\end{align}
This choice of variables also rationalises the root of $\kappa$. The physical region is bounded by the constraints
\begin{align}
    x>0\,,~~~ 0<y<z<1\,,~~~ m^2>0\,.
\end{align}
The symbol alphabet of the MI involves 19 letters (see ref.~\cite{Gehrmann:2015ora} for a detailed description). The results of the MI are computed as Laurent series in $\epsilon$ up to transcendental weight 4 which enables us to get the one-loop polarised amplitudes, ${\cal M}^{[j],(1)}$, to ${\cal O}(\epsilon^2)$ and the two-loop ones, ${\cal M}^{[j],(2)}$ in \eqref{eq:UV-amp-expansion}, to ${\cal O}(\epsilon^0)$ in dimensional regularisation.

While the analytic results of the UV renormalised polarised amplitudes defined in \eqref{eq:HLamps} and \eqref{eq:UV-Amp} are too lengthy to be presented explicitly here, we thus attach these results as ancillary files in Mathematica format along with the \arXiv~submission\footnote{In particular, the file containing the six UV renormalised two-loop amplitudes ${\cal M}^{[j],(2)}$ is about 56 MB.}. Details about conventions and variables of the saved analytic expressions can be found in the {\small ReadMe.txt} submitted along with these files. 
In the following section, we will discuss the universal infrared divergences present in these UV renormalised amplitudes and the (four-dimensional) finite remainders subsequently defined.

\section{IR factorisation and RS Independent Finite Remainders}
\label{sec:irfr}

In a gauge theory like QCD, the UV-finite amplitudes beyond leading order typically contain divergences arising from the soft and collinear configurations of the loop momenta, which appear as poles in dimensional regulator $\epsilon$. Fortunately these IR divergences systematically factor out from the amplitudes to all orders in perturbation theory~\cite{Sen:1982bt,Kidonakis:1998nf}, which  demonstrates a kind of simple universal structure in these additional divergences. The explicit form of these factorised universal IR poles in massless QCD, being only dependent on the nature of the external coloured particles, was first determined up to two-loop order in terms of universal IR subtraction operators in ref.~\cite{Catani:1998bh}.
In ref.~\cite{Sterman:2002qn}, a detailed derivation was presented by exploiting the factorisation and resummation properties of scattering amplitudes which was later generalised to all orders in terms of a soft anomalous dimension matrix in refs.~\cite{Becher:2009cu,Gardi:2009qi}.

\subsection{The RS Independent Finite Remainders of $\mathcal{M}^{[j]}$}
\label{sec:irfr_LP}

As briefly mentioned at the beginning of section~\ref{sec:projectors_LP} and also argued in detail in ref.\cite{Chen:2019wyb}, the bare and UV renormalised polarised amplitudes resulting from the projectors constructed in section~\ref{sec:projectors_LP} are in general different from their counterparts defined in dimensional regularisation schemes such as CDR, HV, DRED and FDH (and hence they should not be compared blindly without conversion).
However, crucially, the properly defined finite remainders in four dimensions are guaranteed to be the same, known as the regularisation-scheme (RS) independence of these objects. 
In other words, the projection prescription adopted here implies a specific regularisation prescription that, despite being different from its companions, remains also \textit{unitary} in the sense as defined in refs.~\cite{vanDamme:1984ig,Catani:1996pk}. 
As long as one is only concerned with quantities that actually contribute to physical observables,  identical results should always be obtained.

To appreciate this crucial point, recall that in this particular prescription, all open Lorentz indices of the (physical) polarisation projectors are set to be D-dimensional and no dimensional splitting is ever introduced, just like in CDR, which consequently ensures the commutation between Lorentz index contraction in the projection and loop integration. This means that applying the projectors defined in \eqref{eq:LPprojectors_canonical} directly to the original Feynman-diagrammatic representation of a loop amplitude should lead to the same polarised amplitudes as would be obtained by applying these projectors to a conceivable (faithful) D-dimensional form factor decomposition representation of that amplitude. The difference between this particular regularisation prescription and the CDR, with $\gamma_5$ regularised and treated in accordance with refs.~\cite{Larin:1991tj,Larin:1993tq}, concerns merely the external particles' state vectors, which are decoupled from the UV and/or IR singularities contained in the loop integration, as is evident from the multiplicative UV renormalisation and IR factorisation as discussed in the preceding sections. Regarding the latter part, captured by certain ``universal'' factors, both regularisation prescriptions are exactly the same. 
Thus the regularisation convention implied by projectors defined in \eqref{eq:LPprojectors_canonical} shares the identical set of renormalisation constants (as given in section~\ref{sec:uv}) and IR anomalous dimensions
as the CDR (with $\gamma_5$ regularised and treated technically in accordance with refs.~\cite{Larin:1991tj,Larin:1993tq}).
From this point of view one can readily expect to end up with the same (four-dimensional) finite remainder in this regularisation prescription as one would obtain from a computation purely within the CDR (and also the same as in any other variant of the dimensional regularisation).
~\\

The IR pole structures in the UV renormalised polarised amplitudes~\eqref{eq:UV-amp-expansion} can be exhibited through a parameterisation in terms of the universal IR subtraction operators  $\mathbf{I}^{(i)}(\epsilon)$, depicted in~\cite{Catani:1998bh} as
\begin{align}
\label{eq:IR-fac}
    &{\cal M}^{[j],(1)} = 2 {\mathbf{I}}^{(1)}(\epsilon) {\cal M}^{[j],(0)} + {\cal M}^{[j],(1)}_{\rm fin}\,,\nonumber\\
    &{\cal M}^{[j],(2)} = 4 {\mathbf{I}}^{(2)}(\epsilon) {\cal M}^{[j],(0)}+2 {\mathbf{I}}^{(1)}(\epsilon) {\cal M}^{[j],(1)} + {\cal M}^{[j],(2)}_{\rm fin}\,.
\end{align}
The explicit expressions of the IR subtraction operators for the current process are given  in the CDR scheme  by
\begin{align}
\label{eq:I1I2}
    &\mathbf{I}^{(1)}(\epsilon)=-C_F\frac{e^{\epsilon\gamma_E}}{\Gamma(1-\epsilon)} \left(\frac{1}{\epsilon^2}+\frac{3}{2\epsilon} \right) \left(-\frac{\mu_R^2}{s}\right)^{\epsilon}\,,\nonumber\\
    &\mathbf{I}^{(2)}(\epsilon)=-\frac{1}{2} \mathbf{I}^{(1)}(\epsilon) \left(\mathbf{I}^{(1)}(\epsilon)+\frac{1}{\epsilon}\beta_0\right) + \frac{e^{-\epsilon\gamma_E}\Gamma(1-2 \epsilon)}{\Gamma(1-\epsilon)} \left(\frac{2}{\epsilon}\beta_0+K\right) \mathbf{I}^{(1)}(2\epsilon) + 2 H^{(2)}(\epsilon)
\end{align}
with the cusp anomalous dimension $K$ and the quantity $H^{(2)}(\epsilon)$~\cite{Becher:2009cu}:
\begin{align}
\label{eq:K-H}
    &K=\left(\frac{67}{18}-\frac{\pi^2}{6}\right) C_A -\frac{5}{9} n_f\,,\nonumber\\
    &H^{(2)}(\epsilon)= - \left( - \frac{\mu_R^2}{s}\right)^{2\epsilon}  \frac{e^{\epsilon\gamma_E}}{\Gamma(1-\epsilon)} \frac{1}{2\epsilon} \Bigg\{C_A C_F \left(-\frac{245}{432}+\frac{23}{16} \zeta_2-\frac{13}{4} \zeta_3\right)\nonumber\\ 
    &~~~~~~~~~~~\qquad\qquad+ C_F^2 \left(\frac{3}{16}-\frac{3}{2} \zeta_2+3 \zeta_3\right) + C_F n_f \left(\frac{25}{216}-\frac{1}{8} \zeta_2\right)\Bigg\}\,.
\end{align}

Based on the discussion given above, we thus expect that the explicit expressions of the IR subtraction operators $\mathbf{I}^{(i)}(\epsilon)$ in CDR given by \eqref{eq:I1I2} can be directly used to predict the IR poles contained in our UV renormalised polarised amplitudes according to Catani's IR factorisation formula \eqref{eq:IR-fac}. 
Indeed this is precisely what we have observed in all our polarised UV renormalised amplitudes, defined by \eqref{eq:UV-Amp} and power expanded in \eqref{eq:UV-amp-expansion}. This thus serves as a strong check on the correctness of our computation. Upon subtracting all these predicted IR singular pieces from the UV-finite polarised amplitudes, we obtain the finite remainders ${\cal M}^{[j],(1)}_{\mathrm{fin}}$ and ${\cal M}^{[j],(2)}_{\mathrm{fin}}$.
The analytic expressions of these finite remainders are too lengthy to be presented explicitly here, but they can be extracted from our UV renormalised polarised amplitudes, attached as ancillary file in Mathematica format, according to \eqref{eq:IR-fac}.

\subsection{Same Finite Remainders Recovered from Form Factor Decomposition}
\label{sec:irfr_FF}

As alluded to in section~\ref{sec:projectors_FF}, we like to cross-check these finite remainders, ${\cal M}^{[j],(1)}_{\mathrm{fin}}$ and ${\cal M}^{[j],(2)}_{\mathrm{fin}}$ obtained using projectors \eqref{eq:LPprojectors_canonical}, by performing also computations based on the conventional form factor decomposition.
In view of the technical differences regarding the UV renormalisations of vector and axial currents as well as the subtle difference regarding linear completeness of the respective decomposition basis in D dimensions, this comparison is conveniently divided into the vector and axial part. 
~\\

Let us start with the vector part of the amplitude, ${\cal M}^{[j]}_{vec}$, where the comparison is really straightforward.
The vector amplitude reconstructed according to \eqref{eq:FFD_vec} with form factors $F_{i,vec}$ projected using \eqref{eq:FFD_vec_projectors} is truly a faithful representation of $\mathcal{M}_{vec}$ in D dimensions (defined by its original Feynman diagrammatic representation), because the basis structures in \eqref{eq:FFD_vec} are linearly complete w.r.t $\bar{v}(p_2) \, \mathbf{\Gamma}^{\mu}_{vec} \, u(p_1)$ in D dimensions regardless of the QCD loop order.
Owing to this, the comparison can actually be made directly for the un-renormalised un-subtracted bare virtual amplitudes, and the agreement should be exact with full D dependence (i.e. no expansion and truncation in $\epsilon$).
Furthermore, since the linear composition with the projections $\mathcal{P}^{\mu}_i \, \bar{v}(p_2) \, \mathbf{\Gamma}_{\mu} \, u(p_1)$ to the polarised amplitudes ${\cal M}^{[j]}$, given in \eqref{eq:HLamps}, is fixed independent of how $\bar{v}(p_2) \, \mathbf{\Gamma}^{\mu} \, u(p_1)$ is computed, it is thus sufficient, and technically more convenient, to compare directly at the level of these projections. 
As already sketched at the end of section~\ref{sec:projectors_FF_vec}, 
the transformation connecting form factors $F_{i,vec}$ to 
$\mathcal{P}^{\mu}_i \, \bar{v}(p_2) \, \mathbf{\Gamma}_{vec,\mu} \, u(p_1)$ 
reads as 
\begin{align} 
\label{eq:fromFFtoLP_vec}
\mathcal{P}^{\mu}_i \, \bar{v}(p_2) \, \mathbf{\Gamma}_{vec, \mu} \, u(p_1) &= F_{1,vec} \, \Big( \mathcal{P}^{\mu}_i \, \bar{v}(p_2) \, u(p_1) \, p_{1,\mu} \Big) 
    \,+\, F_{2,vec} \, \Big( \mathcal{P}^{\mu}_i\, \bar{v}(p_2) \, u(p_1) \, p_{2,\mu} \Big) \nonumber\\
    &\,+\, F_{3,vec} \, \Big( \mathcal{P}^{\mu}_i\, \bar{v}(p_2) \, u(p_1) \, q_{1,\mu} \Big)
    \,+\, F_{4,vec} \, \Big( \mathcal{P}^{\mu}_i\, \bar{v}(p_2) \,\gamma_{\mu} \slashed{q}_1 u(p_1) \Big)\, ,
\end{align}
where $i$ runs from 1 to 6 as in \eqref{eq:LPprojectors_canonical} and the contraction involved in the big parentheses should be done according to D dimensional Lorentz/Dirac algebra. We have checked explicitly that there is an exact agreement for these six bare quantities (with full D dependence) between the direct projection calculation and the one with a detour of conventional D-dimensional form factor decomposition. From this, it follows straightforwardly that the 4-dimensional finite remainders ${\cal M}^{[j]}_{vec,\mathrm{fin}}$ will be the same because the UV renormalisations and IR subtractions proceed identically in both computations.
~\\

Concerning the vector part of the amplitude \eqref{eq:bbZHamplitude} done with a strictly D-dimensional faithful form factor decomposition, there is really no surprise that the same finite remainders are obtained. 
Regarding the axial part, the projectors for projecting our ``axial form factors'' defined in \eqref{eq:FFD_axi_def} are a bit un-conventional due to two points as discussed in section~\ref{sec:projectors_FF_axi}: 
(1) the basis set \eqref{eq:FFD_axi_4} is known to be linearly incomplete in D dimensions for $\mathcal{M}_{axi}$ in question;
(2) all explicit D appearing in the corresponding projectors \eqref{eq:FFD_axial_projectors} are set manually to be 4 even thought it is not an overall D dependence. 
This makes our form factor decomposition for the axial part not qualified as being called D-dimensional faithful.
Still we defined an intermediate axial amplitude $\tilde{\mathcal{M}}_{axi}$ in \eqref{eq:FFD_axi_amp} from the so determined ``axial form factors''. As mentioned before it would be very interesting to see whether the correct 4-dimensional finite remainders ${\cal M}^{[j]}_{axi,\mathrm{fin}}$ could still be obtained in a computation based on such an acrobatic version of axial form factor decomposition.

The comparison can proceed following a similar line as the vector part just discussed above, with one important exception: there is now no more any reason for
expecting that the amplitudes before UV renormalisation and IR subtraction agree, 
and indeed they do not in our computations. Consequently, we first properly renormalise our axial form factors $F_{i,axi}$ in \eqref{eq:FFD_axi_def}
according to section~\ref{sec:uv}, and then apply the previously discussed IR subtractions onto the resulting UV-finite axial form factors
to end up with their finite remainders in four dimensions, denoted as $F_{i,axi,\mathrm{fin}}$ respectively. Afterwards these four-dimensional regular objects are combined together according to 
\begin{align} 
\label{eq:fromFFtoLP_axi}
&\, F_{1,axi,\mathrm{fin}} \, \Big[ \mathcal{P}^{\mu}_i\, \bar{v}(p_2) \,\gamma_5\, u(p_1) \, p_{1,\mu} \Big] 
    \,+\, F_{2,axi,\mathrm{fin}} \, \Big[ \mathcal{P}^{\mu}_i\, \bar{v}(p_2) \,\gamma_5\, u(p_1) \, p_{2,\mu} \Big] \nonumber\\
    &\,+\, F_{3,axi,\mathrm{fin}} \, \Big[ \mathcal{P}^{\mu}_i\, \bar{v}(p_2) \,\gamma_5\, u(p_1) \, q_{1,\mu} \Big]
    \,+\, F_{4,axi,\mathrm{fin}} \, \Big[ \mathcal{P}^{\mu}_i\, \bar{v}(p_2) \,\gamma_{\mu} \gamma_5\,\slashed{q}_1 u(p_1) \Big]\,,
\end{align}
where $i$ runs again from 1 to 6 as in \eqref{eq:LPprojectors_canonical} while the contractions involved in the big square bracket can be done according to four-dimensional Lorentz/Dirac algebra.
We have verified explicitly that these four-dimensional finite remainders composed in \eqref{eq:fromFFtoLP_axi} agree exactly with those from the renormalised and subtracted $\mathcal{P}^{\mu}_i \, \bar{v}(p_2) \, \mathbf{\Gamma}_{axi, \mu} \, u(p_1)$ computed from direct projections as described in section~\ref{sec:projectors_LP}. 
Therefore in this work we confirm that, regarding the process \eqref{eq:process}, 
it is not necessary to construct and use the ``ultimate'' D-dimensional axial decomposition basis, as long as one is only concerned with physical quantities 
(or quantities that actually contribute to physical observables in four dimensions). An acrobatic version of axial form factor decomposition such as described in section~\ref{sec:projectors_FF_axi} is sufficient even in a calculation done with dimensional regularisation.
Of course, a similar statement applies also to the vector part of the amplitude \eqref{eq:bbZHamplitude}, which we also checked explicitly in our computational setup.

\section{Cross Section to NNLO in the Soft-Virtual Approximation}
\label{sec:cssv}

As we discussed in the introduction~\ref{sec:intro}, the dominant mode for the $ZH$ production at LO is the quark-antiquark annihilation which has three channels viz. $s$, $t$ and $u$, as shown in figure~\ref{dia:tree}. The $t,u$ channels result only from $b$ quark annihilation due to the presence of $b$ quark Yukawa coupling. In the SM, the $b$ quark Yukawa coupling is small owing to its proportionality to the $b$ quark mass. Moreover, the distributions of $b$ quark inside the proton is much smaller than those of other light quarks. Consequently, the contributions arising from the $t$- and $u$-channels are expected to be much smaller than that of the $s$-channel. In previous analyses as reviewed in the introduction, contributions from $t,u$-channels are not taken into account due to their expected small size. Nevertheless, it is important to have a definite estimate of the size of these $t,u$-channel contributions for the current precision studies. In this section, we study the numerical impact of the $t,u$-channel contributions. Since these channels do not interfere with the $s$-channel, they can be treated separately. We obtain their contributions up to NLO in QCD using Madgraph \cite{Alwall:2014hca} and to NNLO in the soft-virtual approximation~\cite{Ravindran:2005vv,Ravindran:2006cg,H:2018hqz}, which is also known as the threshold limit. The latter requires the knowledge of the two-loop amplitudes, as well as the tree-level and one-loop amplitudes to high powers in $\epsilon$, which are presented for the first time in this article.~\\

The hadronic cross section of the $ZH$ production that results from the $t,u$-channels of $b$ quark initiated partonic sub-processes is given by
\begin{align}
\label{eq:Xsection}
    \sigma^{ZH}_{tu} = \sum_{a=b,\overline b}\int dx_1 f_{a} (x_1,\mu_F) \int dx_2 {f}_{\bar a}(x_2,\mu_F)  
\sigma^{ZH}_{a \bar a,tu}(x_1,x_2,m_z,m_h,\mu_F)\,.
\end{align}
Here, $x_{1,2}$ are fractions of momenta of incoming hadrons carried away by bottom quarks $b$ and $\bar b$, and ${f}_{a}$ is the parton distribution function (PDF) normalised at the factorisation scale $\mu_F$.  The mass factorised partonic cross section is given by ${\sigma}^{ZH}_{b\bar b,tu}$. 
Both the LO and NLO contributions to $\sigma^{ZH}_{b\bar b,tu}$ are straightforward to obtain. We use Madgraph~\cite{Alwall:2014hca} for this. 
For the NNLO cross section, in addition to the virtual contributions computed in this article, the double-real and real-virtual contributions are also needed. The inclusion of the exact real radiations is beyond the scope of the present work. While, in the absence of this, owing to the universality of soft and collinear contributions, an approximated result of these real radiation corrections can be computed following~\cite{Ravindran:2005vv,Ravindran:2006cg,H:2018hqz}, which combined with the virtual part gives a finite quantity, called the soft-virtual (SV) cross section. At the partonic level, the dominant contribution to the SV cross section comes from the terms proportional to distributions of the kind $\delta(1-z)$ and $D_j(z)$ which is defined as 
\begin{align}
\label{eq:plus-dist}
	D_j(z)\equiv\left(\frac{\log^j(1-z)}{(1-z)} \right)_+\,.
\end{align}
The variable $z$ is defined as $Q^2/s$ where $Q^2=(q_1+q_2)^2$ is the invariant mass square of the final state $ZH$. For the current process, such contributions can arise only from the $b$-quark initiated sub-processes. 
Hence, we write the hadronic cross section resulting from $t,u$ channels as
\begin{eqnarray}
\sigma_{b\overline b,tu}^{ZH} = \sigma_{b \overline b,tu}^{ZH,\text{SV}} + \sigma_{b \overline b,tu}^{ZH,\text{Hard}}\,.
\end{eqnarray}
Thanks to the factorisation of the matrix elements as well as the phase space into 
soft and hard parts in the soft limit, 
the SV part ($\sigma_{b \overline b,tu}^{ZH,\text{SV}}$) of the inclusive cross section 
is found to be
\begin{align}
\label{eq:Xsection-SV-Hadron}
    \sigma^{ZH,\text{SV}}_{b \overline b,tu} =& \int {d Q^2 \over Q^2} 
\sum_{b,\overline b} \int dx_1 f_b(x_1,\mu_F) \int dx_2 f_{\overline b} (x_2,\mu_F)  
{1 \over 2  s} \prod_{n=1}^2 \int d \phi(q_n) 
\nonumber\\
&
\times (2 \pi)^D \delta^D\Big(p_1+p_2-\sum_{n=1}^2 q_n\Big)
\overline \sum \Delta_{b\overline b}^{ZH,\text{SV}} 
\left(\{p_j\cdot q_k\},z,Q^2,\mu_F\right)\,.
\end{align}
where $\Delta_{b\overline b}^{ZH,\text{SV}}$ is the mass factorised partonic cross section in the threshold limit which
is calculable order by order in strong coupling constant $a_s$. Suppressing the obvious arguments, we expand 
$\Delta_{b\overline b}^{ZH,\text{SV}}$ as
\begin{align}
\label{eq:partonic-Xsection-expand}
    \Delta_{b\overline b}^{ZH,\text{SV}}=\sum_{j=0}^{\infty} a_s^j(\mu_R^2) \Delta_{b\overline b}^{\text{SV},(j)}(\mu_R^2)
\,.
\end{align}
Using the one and two loop amplitudes computed in this article, the universal soft distribution functions 
and the Altarelli-Parisi kernels, $\Delta_{b\overline b}^{\text{SV},(j)}$ are obtained
in terms of cusp $A_q$, soft $f_q$ and collinear $B_q$ anomalous dimensions.
Setting the renormalisation and factorisation scales at $Q$, i.e., $\mu_R=\mu_F=Q$,
we find
\begin{eqnarray}
\label{eq:SV-Analytical-Res}
   \Delta^{\text{SV},(0)}_{b \overline b} &=&
   \delta(1-z)  |{\cal M}^{(0)}_{0}|^2\,,
\nonumber\\
   \Delta^{\text{SV},(1)}_{b \overline b} &=&
    \delta(1-z) 
    \Bigg[ 2 \Re\Big({\cal M}^{(0)}_{2}   {\cal M}^{\star(1)}_{-2}
        +  {\cal M}^{(0)}_{1}   {\cal M}^{\star(1)}_{-1} + {\cal M}^{(0)}_{0}   {\cal M}^{\star(1)}_{0}\Big ) 
	+ \big( - 2 f_1^{q} - 4 B_1^{q} \big)
\nonumber\\&&
\times        2 \Re \Big({\cal M}^{(0)}_{1}   {\cal M}^{\star(0)}_{0}\Big)
	+ 4 A_1^{q} \Big( |{\cal M}^{(0)}_{1}|^2 + 2 \Re \Big( {\cal M}^{(0)}_{0}   {\cal M}^{\star(0)}_{2} \Big)\Big)
        + 2 {\cal \overline G}_1^{q,(1)} |{\cal M}^{(0)}_{0}|^2
	\Bigg]
\nonumber\\&&
	+ {\cal D}_0  |{\cal M}^{(0)}_{0}|^2    \Bigg[  - 2  f_1^{q}\Bigg]
	+ {\cal D}_1  |{\cal M}^{(0)}_{0}|^2    \Bigg[ 4  A_1^{q}  \Bigg]\,,
\nonumber\\
\Delta^{\text{SV},(2)}_{i,b \overline b}   &=& 
	\delta(1-z) \Bigg[
        |{\cal M}^{(1)}_{0}|^2 
	+ 2 \Re\Big({\cal M}^{(1)}_{2}   {\cal M}^{\star(1)}_{-2} 
        +  {\cal M}^{(1)}_{1}   {\cal M}^{\star(1)}_{-1} + {\cal M}^{(0)}_{4}   {\cal M}^{\star(2)}_{-4}
        + {\cal M}^{(2)}_{-3}   {\cal M}^{\star(0)}_{3} 
\nonumber\\&&
      + {\cal M}^{(2)}_{-2}   {\cal M}^{\star(0)}_{2}
        + {\cal M}^{(2)}_{-1}   {\cal M}^{\star(0)}_{1} +    {\cal M}^{(2)}_{0}   {\cal M}^{\star(0)}_{0}\Big )
	+ \big( - f_2^{q} - 2 B_2^{q} - 4 {\cal \overline G}_1^{q,(1)} f_1^{q} 
\nonumber\\&&
        - 8 {\cal \overline G}_1^{q,(1)} B_1^{q}\big) 
	2 \Re \Big({\cal M}^{(0)}_{1}   {\cal M}^{\star(0)}_{0}\Big)
	+\big(-2f_1^{q}-4 B_1^{q}\big) \Big( \beta_0 |{\cal M}^{(0)}_{1}|^2 + 2 \Re \Big( {\cal M}^{(1)}_{-2}   
          {\cal M}^{\star(0)}_{3}
\nonumber\\&&
	+ {\cal M}^{(1)}_{-1}   {\cal M}^{\star(0)}_{2} 
	+ {\cal M}^{(0)}_{1}   {\cal M}^{\star(1)}_{0}
	+ {\cal M}^{(0)}_{0}   {\cal M}^{\star(1)}_{1}
	+ \beta_0 {\cal M}^{(0)}_{0}   {\cal M}^{\star(0)}_{2}\Big) \Big)
	+ 2 \big(f_1^{q}\big)^2 \Big( |{\cal M}^{(0)}_{1}|^2 
\nonumber\\&&
        - \zeta_2 |{\cal M}^{(0)}_{0}|^2 + 2 \Re \Big( {\cal M}^{(0)}_{0}   {\cal M}^{\star(0)}_{2} \Big)  \Big )
	+ \big(8 B_1^{q}f_1^{q} + 8 (B_1^{q})^2 + A_2^{q}\big) \Big( |{\cal M}^{(0)}_{1}|^2 
\nonumber\\&&
       + 2 \Re \Big( {\cal M}^{(0)}_{0}   {\cal M}^{\star(0)}_{2} \Big)\Big)
	+ \big( A_1^{q} \big)2 \Re\Big( 4 {\cal M}^{(1)}_{-2}   {\cal M}^{\star(0)}_{4}
       + 4 {\cal M}^{(1)}_{-1}   {\cal M}^{\star(0)}_{3}
       + 4 {\cal M}^{(1)}_{0}   {\cal M}^{\star(0)}_{2}
\nonumber\\&&
       + 4 {\cal M}^{(1)}_{1}   {\cal M}^{\star(0)}_{1}
       + 4 {\cal M}^{(1)}_{2}   {\cal M}^{\star(0)}_{0}
       + 6 \beta_0 {\cal M}^{(0)}_{1}   {\cal M}^{\star(0)}_{2}
	+ 6 \beta_0 {\cal M}^{(0)}_{0}   {\cal M}^{\star(0)}_{3}\Big)
\nonumber\\&&
	+ \big(- 8 A_1^{q} f_1^{q} \big) \Big( 2 \Re \Big( {\cal M}^{(0)}_{1}   {\cal M}^{\star(0)}_{2}
	+ {\cal M}^{(0)}_{0}   {\cal M}^{\star(0)}_{3}\Big) + \zeta_{3} |{\cal M}^{(0)}_{0}|^2 \Big)
	+ \big(- 16 A_1^{q} B_1^{q} \big) 
\nonumber\\&&
\times 2 \Re \Big( {\cal M}^{(0)}_{1}   {\cal M}^{\star(0)}_{2}
	+ {\cal M}^{(0)}_{0}   {\cal M}^{\star(0)}_{3}\Big) 
	+ \big(8 (A_1^{q})^2 \big) \Big( |{\cal M}^{(0)}_{2}|^2 - \frac{1}{10}\zeta_{2}^2 |{\cal M}^{(0)}_{0}|^2
\nonumber\\&&
	+ 2\Re \Big( {\cal M}^{(0)}_{1}   {\cal M}^{\star(0)}_{3} + {\cal M}^{(0)}_{0}   {\cal M}^{\star(0)}_{4}
	\Big) \Big)
	+ \big( 2  {\cal \overline G}_1^{q,(1)} \big) 2 \Re \Big( {\cal M}^{(0)}_{2} {\cal M}^{\star(1)}_{-2}
	+ {\cal M}^{(0)}_{1}   {\cal M}^{\star(1)}_{-1} 
\nonumber\\&&
        + {\cal M}^{(0)}_{0}   {\cal M}^{\star(1)}_{0} \Big)
	+ \big({8 \cal \overline G}_1^{q,(1)} A_1^{q}\big) \Big( |{\cal M}^{(0)}_{1}|^2 + 2 \Re \Big(  {\cal M}^{(0)}_{0}   {\cal M}^{\star(0)}_{2} \Big)   \Big)  
	+ \Big(2 \big( {\cal \overline G}_1^{q,(1)} \big)^2 
\nonumber\\&&
     + 2 \beta_0 {\cal \overline G}_2^{q,(1)} + {\cal \overline G}_1^{q,(2)}\Big) \Big( |{\cal M}^{(0)}_{0}|^2 \Big)\Bigg]
	+ {\cal D}_0 \Bigg[  \Big(16\zeta_3 (A_1^q)^2 - 4  {\cal \overline G}_1^{q,(1)} f_1^{q}  -4 \beta_0  
       {\cal \overline G}_1^{q,(1)} 
\nonumber\\&&
       - 2 f_2^{q}\Big) \Big( |{\cal M}^{(0)}_{0}|^2 \Big)
	+ \Big( -2 f_1^{q}\Big) \Big( 2 \Re \Big( {\cal M}^{(0)}_{2} {\cal M}^{\star(1)}_{-2}
        + {\cal M}^{(0)}_{1}   {\cal M}^{\star(1)}_{-1} + {\cal M}^{(0)}_{0}   {\cal M}^{\star(1)}_{0} \Big) 
\nonumber\\&&
	 + \Big( 4 \big(f_1^{q})^2 + 8 B_1^{q}f_1^{q} 
\Big)
        2 \Re \Big({\cal M}^{(0)}_{1}   {\cal M}^{\star(0)}_{0}\Big) 
	+ \big( -8 A_1^q f_1^q\big)\Big( |{\cal M}^{(0)}_{1}|^2 - \zeta_2 |{\cal M}^{(0)}_{0}|^2 
\nonumber\\&&
       + 2 \Re \Big( {\cal M}^{(0)}_{0}   {\cal M}^{\star(0)}_{2} \Big)  \Big ) \Bigg]
	+ {\cal D}_1 \Bigg[ \Big(8 {\cal \overline G}_1^{q,(1)} A_1^{q} + 4 \beta_0 f_1^{q}  + 4 (f_1^q)^2 + 4 A_2^{q}\Big) \Big( |{\cal M}^{(0)}_{0}|^2 \Big)
\nonumber\\&&
	+ \Big( 4 A_1^{q}\Big) \Big( 2 \Re \Big( {\cal M}^{(0)}_{2} {\cal M}^{\star(1)}_{-2}
        + {\cal M}^{(0)}_{1}   {\cal M}^{\star(1)}_{-1} + {\cal M}^{(0)}_{0}   {\cal M}^{\star(1)}_{0} \Big)
	+ \Big( -8 f_1^{q} A_1^{q} 
\nonumber\\&&
        - 16 B_1^{q}A_1^{q} 
\Big)
        2 \Re \Big({\cal M}^{(0)}_{1}   {\cal M}^{\star(0)}_{0}\Big)
	+ \Big(16 (A_1^{q})^2\Big)\Big( |{\cal M}^{(0)}_{1}|^2 - \zeta_2 |{\cal M}^{(0)}_{0}|^2 
\nonumber\\&&
        + 2 \Re \Big( {\cal M}^{(0)}_{0}   {\cal M}^{\star(0)}_{2} \Big)  \Big ) \Bigg]
       + {\cal D}_2      \Bigg[ \Big( - 12  A_1^{q} f_1^{q}  - 4 \beta_0  A_1^{q}\Big) |{\cal M}^{(0)}_{0}|^2  \Bigg]
\nonumber\\&&
	+ {\cal D}_3  |{\cal M}^{(0)}_{0}|^2    \Bigg[ 8  (A_1^{q})^2 \Bigg]\,.
\end{eqnarray}
Here $\Re$ denotes the real part of the quantity within the respective parenthesis. The quantity ${\cal M}^{(j)}_k$ is the coefficient of $a_s^j\, \epsilon^k$ in the expansion of the matrix element ${\cal M}$ in \eqref{eq:reln-mat-polMat} through:
\begin{align}
\label{eq:expand-matrix-element}
    {\cal M}= \sum_{j=0}^{\infty} a_s^j(\mu_R^2) {\cal M}^{(j)} = \sum_{j=0}^{\infty} \sum_{k=-2j}^{\infty} a_s^j(\mu_R^2) \epsilon^k {\cal M}^{(j)}_k\,.
\end{align}
Using the relation between the unpolarised squared matrix element squares and the squared modulus of the polarised amplitudes given in \eqref{eq:sqM}, we relate these quantities order by order in $a_s$ which read
\begin{align}
\label{eq:ren-matsq-polmatsq}
    &\left|{\cal M}^{(0)}\right|^2=\sum_{j=1}^6 \left|{\cal M}^{[j],(0)}\right|^2\,,\nonumber\\
    &2 \Re \left({\cal M}^{\star(0)} {\cal M}^{(1)}\right)=\sum_{j=1}^6 2 \Re \left({\cal M}^{\star[j],(0)} {\cal M}^{[j],(1)}\right)\,,\nonumber\\
    &2 \Re \left({\cal M}^{\star(0)} {\cal M}^{(2)}\right)+\left|{\cal M}^{(1)}\right|^2=\sum_{j=1}^6 \Big\{ 2 \Re \left({\cal M}^{\star[j],(0)} {\cal M}^{[j],(1)}\right) +\left|{\cal M}^{[j],(1)}\right|^2 \Big\}\, ,
\end{align}
where the polarised amplitudes are expanded in powers of $a_s$ in \eqref{eq:UV-amp-expansion}. Using 
the results of the polarised amplitudes which are computed in this article to ${\cal O}(\epsilon^4)$, 
${\cal O}(\epsilon^2)$ and ${\cal O}(\epsilon^0)$ at tree, one- and two-loop level, respectively, 
we have obtained the square of the matrix elements up to two-loop level. 
Moreover, by expanding the square of the matrix elements that appeared on the left-hand side of the aforementioned equation in powers of $\epsilon$ following \eqref{eq:expand-matrix-element} and comparing these with the explicit results obtained in this article, we have extracted the relevant quantities appearing in the results of the SV cross section in \eqref{eq:SV-Analytical-Res}.

The cusp~\cite{Vogt:2000ci,Berger:2002sv,Moch:2004pa,Moch:2005tm,Becher:2009qa,Baikov:2009bg, Gehrmann:2010ue}, soft, and collinear anomalous dimensions are given by
\begin{align}
\label{eq:ano-dim}
 A^q_1 &= 4 C_F \,,\nonumber \\
 A^q_2 &= 8 C_F C_A \Big( \frac{67}{18} - \zeta_2 \Big) + 8 C_F n_f \Big( -\frac{5}{9} \Big) \,,\nonumber \\
 B^q_1 &= 3 C_F \,,\nonumber\\
 B^q_2 &= \frac{1}{2} \Big \{ C_F^2 \big (3-24\zeta_2+48\zeta_3 \big)
        + C_AC_F\Big(\frac{17}{3} + \frac{88}{3}\zeta_2 -24 \zeta_3  \Big)
+ C_Fn_fT_F\Big(-\frac{4}{3} -\frac{32}{3}\zeta_2 \Big) \Big\} \,, \nonumber \\
 f_1^q &= 0 \,,\nonumber \\
 f_2^q &= C_A C_F \Big( -\frac{22}{3} {\zeta_2} - 28 {\zeta_3} + \frac{808}{27} \Big)
        + C_F n_f \Big(\frac{4}{3} {\zeta_2} - \frac{112}{27} \Big) \,.
\end{align} 
The universal constants ${\overline {\cal G}}^{q,(j)}_k$ arising from the soft-collinear distribution 
are given by~\cite{Ravindran:2006cg}
\begin{align}
\label{eq:curlyG}
  {\overline {\cal G}}^{q,(1)}_1 &= C_F \left( - 3 \zeta_2 \right) \,,\nonumber \\
  {\overline {\cal G}}^{q,(1)}_2 &= C_F \left( \frac{7}{3} \zeta_3 \right) \,, \nonumber\\
  {\overline {\cal G}}^{q,(2)}_1 &=  C_F n_f  \left( - \frac{328}{81} + \frac{70}{9} \zeta_2 + \frac{32}{3} \zeta_3 \right)
             + C_A C_F  \left( \frac{2428}{81} - \frac{469}{9} \zeta_2
                       + 4 {\zeta_2}^2 - \frac{176}{3} \zeta_3 \right) 
\end{align}
where, $\zeta_2=1.64493407\cdots,\zeta_3=1.20205690\cdots$. Now using the matrix elements along with other universal quantities in \eqref{eq:SV-Analytical-Res}, we 
obtain the SV cross section up to NNLO which is finite in the limit $\epsilon\rightarrow 0$. 
The cancellation of soft, collinear singularities among virtual, real emissions and mass factorisation
kernels in the SV cross section serves as a check of the correctness of our 
renormalised polarised amplitudes, despite that they are given for a specific regularisation prescription 
implied by projectors devised in section~\ref{sec:projectors_LP}.
The size of the partonic SV cross section at NLO and NNLO are 16 KB and 29 MB respectively, hence we skip presenting them explicitly. Instead the results are included in Mathematica format as ancillary files 
along with the \arXiv~submission.~\\

In the rest of this section, we discuss the numerical result of the total cross section of $ZH$ production at the LHC. The center-of-mass energy is taken to be ${\sqrt S}=13$ TeV. We use MMHT2014 PDF sets obtained through LHAPDF interface \cite{Buckley:2014ana}, and use the five-flavour scheme throughout. The renormalisation and factorisation scales are set at $\mu_R=\mu_F=m_h + m_z$. 
The Fermi constant is $G_F = 1.16637\times 10^{-5}$ GeV$^{-2}$. The width of $Z$ is given by $\Gamma_Z = 2.4952$ GeV. We use running $b$ quark mass in the Yukawa coupling. We choose $m_b(m_b)=4.18$ GeV and evolve it to appropriate scales using the renormalisation group equation.
In table~\ref{tab:Xsection}, 
\begin{table}[h!]
\centering
    \begin{tabular}{|c|c|c|c|c|c|} 
      \hline
    Order  &           $s$-channel & EW          &  $ \sigma^{ZH}_{gg}$ &   $\sigma^{ZH}_{q\bar q}$(top) & $(t+u)$-channel\\
     \hline \hline 
       LO  &           5.897 10$^{-1}$    &           -3.111 10$^{-2}$   &       -         &        -         & 2.989 10$^{-4}$\\ 
       \cline{1-6}
       NLO &           7.756 10$^{-1}$    &              -               &       -         &        -         & 2.934 10$^{-4}$\\
       \cline{1-6}
       NNLO&           8.015 10$^{-1}$    &              -               & 5.051 10$^{-2}$ & 9.442 10$^{-3}$  & 3.027 10$^{-4}$\\
       \cline{1-6}
       N$^3$LO$_{\rm SV}$& 8.013 10$^{-1}$    &              -               &      -          &        -         &   -             \\
       \hline 
    \end{tabular}
    \caption{QCD and electroweak (EW) contributions to $ZH$ production at the LHC with 
center of mass energy 13 TeV. The $\sigma^{ZH}_{gg}$ refers to the contribution coming from the gluon initiated sub-processes. The top quark loop contribution is denoted by $\sigma^{ZH}_{q\bar q}$(top). The $(t+u)$-channel contribution at NNLO is under the SV approximation. All the numbers of the cross-sections are in unit pb.}
      \label{tab:Xsection}
\end{table}
we present QCD (and electroweak) corrections to the $ZH$ production at the LHC$@$13TeV at various orders, with contributions from $t,u$-channels separated from that of the $s$-channel. The $s$-channel contributions include only those processes where the Higgs boson couples to $Z$, whereas the $(t+u)$-channel contributions contain those where the Higgs boson is radiated from the $b$ quark.
The $s$-channel contributions are obtained using the program {\tt vh@nnlo}~\cite{Brein:2012ne} where the electroweak (EW) corrections~\cite{Ciccolini:2003jy, Brein:2004ue} are also implemented. 
The NLO QCD correction increases the total inclusive cross section of the $ZH$ production at the LHC@13TeV by 31$\%$, and the NNLO Drell-Yan-like correction alone by another 4$\%$, and the threshold contribution at N$^3$LO gives less than 1$\%$. Apart from the Drell-Yan like contributions, starting from NNLO in QCD there appear also sub-processes induced by the top quark loop from which the Higgs is radiated. In addition, there is also the gluon initiated sub-processes at NNLO which give rise to about 6$\%$ correction for $ZH$ production. (See table~2 in ref.~\cite{Kumar:2014uwa} for the sizes of various contributions to $ZH$ production at the LHC at different center-of-mass energies.) Regarding the contributions from the $t,u$-channels, we obtain the result up to NLO in QCD using Madgraph \cite{Alwall:2014hca}, and compute the hadronic cross section (\ref{eq:Xsection-SV-Hadron}) at the NNLO using an in-house fortran routine that employs \ginac~\cite{Bauer:2000cp} for the numerical evaluation of polylogarithms~\cite{Vollinga:2004sn}. 
From the numbers in the last column of table~\ref{tab:Xsection}, we see that these $t,u$-channels give a three orders of magnitude smaller contribution compared to that of the $s$-channel one.

\section{Conclusions}
\label{sec:con}

A precise knowledge of the  $ZH$ process is important for fully exploring Higgs physics at hadron colliders, especially to meet the foreseeable precision requirements from future collider experiments for detailed studies of the 125 GeV Higgs boson (as well as potentially non-standard Higgs bosons). Through the work presented in this article we provide the analytic results of the two-loop massless QCD corrections to the $b$-quark-induced $ZH$ process that involves a non-vanishing $b$-quark Yukawa coupling $\lambda_b$, which is a necessary ingredient of the complete  $\mathcal{O}(\alpha_s^2)$ QCD corrections to this process in the five-flavour scheme.

The computation was performed by projecting the D-dimensional scattering amplitudes (in their Feynman-diagrammatic representations) directly onto an appropriate set of Lorentz structures related to the linear polarisation states of the $Z$ boson, as explained in detail in section~\ref{sec:projectors_LP}. 
We emphasize that the renormalised polarised amplitudes, attached as ancillary files, are given for a specific regularisation prescription implied by these special projectors in use, and hence they should not be compared blindly with their counterparts defined in various other DR schemes (e.g.~CDR, HV, DRED and FDH, etc). However, the crucial point is that this special dimensional regularisation prescription is also unitary as argued in ref.\cite{Chen:2019wyb}, and more importantly it shares the identical set of renormalisation constants and IR anomalous dimensions as the CDR (with $\gamma_5$-related objects treated technically in accordance with refs.~\cite{Larin:1991tj,Larin:1993tq}).
Because of this, one will always end up with the same properly defined RS-independent finite remainders, which is eventually required for computing physical observables.
We have verified explicitly that the IR poles contained in our renormalised amplitudes match with those predicted by Catani's IR factorisation formula in section~\ref{sec:irfr} as well as those according to the soft-virtual factorisation formula in section~\ref{sec:cssv}. 
This serves as a strong check of the correctness of our results.

In addition, regarding the vector part of the amplitude \eqref{eq:bbZHamplitude} considered in this publication, we have cross-checked the finite remainders of these renormalised polarised amplitudes, defined according to \eqref{eq:IR-fac},
with those obtained from a strictly D-dimensional conventional form factor decomposition approach and found exact agreement.  
Concerning the conventional form factor decomposition of amplitudes involving axial currents regularised in D dimensions (with a non-anticommuting $\gamma_5$ prescription like \eqref{eq:gamma5}), we have dedicated a few specialised discussions to an interesting issue resembling a bit a similar one in $q \bar{q} \rightarrow Q\bar{Q}$~\cite{Glover:2004si,Abreu:2018jgq}, albeit of a different technical origin.  Despite all concerns discussed in section~\ref{sec:projectors_FF_axi}, through computations accomplished in this article we confirm that regarding the process \eqref{eq:process} it is not necessary to construct and use the ``ultimate'' D-dimensional axial decomposition basis. An acrobatic version of axial form factor decomposition such as described in section~\ref{sec:projectors_FF_axi} is sufficient even for a calculation done with dimensional regularisation (as long as one is only concerned with physical quantities in four dimensions). And a similar statement applies also to the vector part of the amplitude \eqref{eq:bbZHamplitude} covered in our computations.  We think this observation should also hold in cases beyond the one considered in this article. This conclusion has the potential to simplify multi-loop  calculations, particularly those involving axial coupling regularised in D dimensions.

For a quantitative estimate of the size of the non-Drell-Yan contributions, we have studied their 
numerical impact on the total cross section of $ZH$ production in section~\ref{eq:Xsection}. We find 
that these non-Drell-Yan processes give about three orders of magnitude smaller contributions compared 
to that of the $s$-channel Drell-Yan-like processes. Nevertheless, the availability of these polarised amplitudes for the process \eqref{eq:process} would allow us to combine the $b$-quark-induced $ZH$ production with the subsequent decay of the $Z$ boson with full spin correlations accounted for.

\section*{Acknowledgements}

We thank G. Das, S. Borowka, C. Duhr, M.~C. Kumar and T. Gehrmann for their help in numerical analysis. We are grateful to W.~Bernreuther for comments on the manuscript. A.~H.~A. and P.~M. thank V. Vaibhav and G. Sood for consultation, IMSc Admin Staff G. Srinivasan for helping with the cluster facility. The Feynman diagrams are generated using \xml~\cite{nicolas_deutschmann_2016_164393}.

\appendix
\label{sec:appendix}

\section{Results of Linearly Polarised $b\bar b ZH$ Amplitudes at the Tree Level}
\label{secA:LO-results}

For the readers' convenience as well as to be explicit about the electroweak coupling factors suppressed in the ancillary files attached, we document here the complete expressions of projections at the tree level using the linearly polarised projectors introduced in~\eqref{eq:LPprojectors_canonical}:
\begin{align}
\label{eq:LO-LP-Amp}
    \mathcal{P}_1^{\mu}\; \bar v(p_2) \mathbf{\Gamma}_{\mu} u(p_1) =& i \delta_{{\bar j_2}{j_1}} v_c \frac{s}{tu} (t+u) \Big(-2 m_z^6+t u (2 s + t + u)+m_z^4 \left(4 s + 3 (t + u)\right) \nonumber\\
    &-m_z^2 \left(2 s^2 + t^2 + 4 t u + u^2 + 3 s (t + u)\right) \Big)\,,\nonumber\\
    \mathcal{P}_2^{\mu}\; \bar v(p_2) \mathbf{\Gamma}_{\mu} u(p_1) =& - \delta_{{\bar j_2}{j_1}} a_c \frac{s}{2tu} (D-3) (t-u) \Big(m_z^4 + t u - m_z^2 (s + t + u)\Big)\,,\nonumber\\
    \mathcal{P}_3^{\mu}\; \bar v(p_2) \mathbf{\Gamma}_{\mu} u(p_1) =& -i \delta_{{\bar j_2}{j_1}} v_c \frac{2s}{tu} m_z^2 (t-u)   (m_z^2 - s - t - u)\,,\nonumber\\
    \mathcal{P}_4^{\mu}\; \bar v(p_2) \mathbf{\Gamma}_{\mu} u(p_1) =& - i \delta_{{\bar j_2}{j_1}} a_c \frac{s}{6tu} (t-u) (6 - 5 D + D^2) \left(6 m_z^2 - 6 s - (-1 + n) (t + u)\right)\nonumber\\
    &\times \left(m_z^4 + t u - m_z^2 (s + t + u)\right)\,,\nonumber\\
    \mathcal{P}_5^{\mu}\; \bar v(p_2) \mathbf{\Gamma}_{\mu} u(p_1) =&  \delta_{{\bar j_2}{j_1}} v_c \frac{s}{2tu} (D-3)  (t + u) \left(m_z^4 + t u - m_z^2 (s + t + u)\right)\,,\nonumber\\
    \mathcal{P}_6^{\mu}\; \bar v(p_2) \mathbf{\Gamma}_{\mu} u(p_1) =& - i \delta_{{\bar j_2}{j_1}} a_c \frac{1}{6tu} (6 - 5 D + D^2) (t + u) \Big(-2 m_z^2 (D-4) + 6 s\nonumber\\
    &+ (D-4) (t + u)\Big) \Big(m_z^4 + t u - m_z^2 (s + t + u)\Big)
\end{align}
with 
\begin{align}
\label{eq:vc-ac}
    &v_c = \frac{e^2 m_b m_z (-4 m_w^2 + m_z^2)}{24 m_w^2 (m_w^2 - m_z^2)}\,,\nonumber\\
    &a_c = -\frac{e^2 m_b m_z^3}{8 m_w^2 ( m_w^2 -  m_z^2)}\,,\nonumber\\
    &e = \sqrt{4\pi \alpha}\, ,
\end{align}
where for completeness we have also restored explicitly the color structure $\delta_{{\bar j_2}{j_1}}$.
The vector and axial couplings are denoted by $v_c$ and $a_c$, respectively, and
$m_w$ and $m_b$ are the masses of massive $W$-boson and $b$-quark. 
The non-zero mass of $b$-quark is introduced through the $b$-quark Yukawa coupling. The strength of the electromagnetic interaction is encapsulated through the fine-structure constant $\alpha$. 
The ancillary file, named ``{\small Linearly\_Polarised\_Partial\_Amplitudes\_Tree.m}'', contains the expressions in \eqref{eq:LO-LP-Amp} up to the (overall) color structure $\delta_{{\bar j_2}{j_1}}$ and the clearly separated $v_c$ and $a_c$ couplings which are all suppressed for simplicity. 
The polarised amplitudes defined in \eqref{eq:HLamps} at the tree level can be subsequently composed using these projections given in \eqref{eq:LO-LP-Amp}. Since we work to the leading order in $\alpha$, this convention about electroweak coupling factors applies also to our UV renormalised amplitudes provided in ancillary files attached along with the \arXiv~submission.

\section{Results of Leading Order Partonic Cross Section}
\label{secA:LO-Xsection-results}

Here we present the leading order partonic cross section resulting from the $t$ and $u$ channels:
\begin{align}
\label{eq:LO-Xsection}
\Delta^{\rm SV,(0)}_{b\bar b} =  \delta(1-z)\Big\{&a_c^2\frac{4}{m_z^2 t^2 u^2}\Big( m_z^6 (t-u)^2 + 2 s t^2 u^2 + m_z^2 t u (-2 s^2 + (t - u)^2 \nonumber\\
&- 2 s (t+u)) -m_z^4 (( t-u)^2(t+u) +s(t^2 - 4 t u + u^2))\Big)\nonumber\\
&+v_c^2\frac{4}{t^2 u^2}\Big( m_z^4 (t+u)^2 + t u (2 s^2 + 2 s (t+u) + (t+u)^2) \nonumber\\&- m_z^2 ((t+u)^3 + s (t^2 +4 t u + u^2))\Big) \Big\}\,.
\end{align}

\bibliography{bbzh} 
\bibliographystyle{JHEP}
\end{document}